\documentclass[10pt,journal,compsoc]{IEEEtran}

\newif\iffast
\fasttrue

\ifCLASSOPTIONcompsoc
  \usepackage[nocompress]{cite}
\else
  \usepackage{cite}
\fi

\usepackage{times}
\usepackage{booktabs}
\usepackage{colortbl}
\usepackage{dblfloatfix}
\usepackage{xfrac}

\usepackage{multirow}
\usepackage{lipsum}
\usepackage{multicol}
\usepackage[utf8]{inputenc} 
\usepackage[T1]{fontenc}
\newcommand{\para}[1]{\vspace{0.2cm}\noindent\textbf{\textsf{#1}}\qquad}
\usepackage[cmyk]{xcolor}
\usepackage{url}
\usepackage[numbers]{natbib}  

\usepackage{amsmath,amssymb,amsfonts}
\usepackage{algorithmic}
\usepackage{textcomp}
\usepackage{xcolor}
\usepackage{soul}

\newcommand{\eg}{\textit{e.g.}}
\newcommand{\ie}{\textit{i.e.}}
\newcommand{\etal}{\textit{et al.}}

\newcommand{\hide}[1]{}

\newcommand{\greenpos}[1]{\textcolor{green}{#1}}
\newcommand{\redneg}[1]{\textcolor{red}{#1}}

\def\BibTeX{{\rm B\kern-.05em{\sc i\kern-.025em b}\kern-.08em
    T\kern-.1667em\lower.7ex\hbox{E}\kern-.125emX}}

\usepackage{pgfplots}    
\usepgfplotslibrary{groupplots}
\pgfplotsset{compat=1.17}
\selectcolormodel{cmyk}
\usepgfplotslibrary{fillbetween, colorbrewer, colormaps}
\usepackage{pgfplotstable}
\usepackage{tikz}    
    
\usetikzlibrary{shapes,arrows,spy}
\usetikzlibrary{calc, plotmarks, backgrounds}
\usetikzlibrary{decorations.pathreplacing,decorations.markings,arrows}
\usetikzlibrary{arrows,decorations.markings, shapes}
\usetikzlibrary{shapes.arrows}
\usetikzlibrary{patterns, hobby, fit, positioning, automata, arrows.meta}

\tikzstyle{textnode} = [rectangle, inner sep=0pt,outer sep=0,execute at begin node={\strut}, font=\small]  
\tikzstyle{bnode} = [circle, draw, fill=black, minimum size=4mm, text=white, outer sep=1.5pt]  
\tikzstyle{enode} = [circle, thick, draw, fill=gray!20, minimum size=8mm, inner sep=0.25pt, outer sep=1.5pt]  

\tikzstyle{inode} = [circle, thick, draw, minimum size=1.5mm, outer sep=1.25pt]  
\tikzstyle{nt} = [draw, inner xsep=1.5, fill=gray!5, minimum size=3mm, minimum size=1mm, outer sep=1.5pt]  
\tikzstyle{tnode} = [minimum size=5mm, font=\large]  
\tikzstyle{hnode} = [enode, very thick, draw=blue, outer sep=1.5pt]  
\tikzstyle{hidden} = [draw=none]
\tikzstyle{edge} = [->, thick, >=stealth',  auto,]  
\tikzstyle{edge} = [auto,]  
\tikzstyle{iedge} = [edge, ultra thick, draw=blue]   
\tikzstyle{bedge} = [edge, draw=red]  
\tikzstyle{faded} = [opacity=0.60, text opacity=0.60]
\tikzstyle{Arrow} = [line width=0.5mm, draw=red!80, fill=red!80, -{Triangle[length=1.5mm,width=1.25mm]}]
\tikzstyle{highlight} = [fill=gray!50, very thick]
\tikzstyle{shrink} = [scale=0.5, transform shape]
\tikzstyle{treenode} = [enode, shrink]
\tikzstyle{diredge} = [-{latex'[scale=0.5]}]

\definecolor{blue1}{HTML}{c2ebf7}
\definecolor{blue2}{HTML}{98baca}
\definecolor{blue3}{HTML}{718c9e}
\definecolor{blue4}{HTML}{4d6073}
\definecolor{blue5}{HTML}{2c3849}
\definecolor{blue6}{HTML}{0e1323}
\definecolor{blue7}{HTML}{0e1323}
\definecolor{kron_blue}{HTML}{6495ED}
\definecolor{netgan_green}{HTML}{32CD32}
\definecolor{linae_red}{HTML}{FF4500}
\definecolor{rnn_blue}{HTML}{B0C4DE}
\definecolor{navy}{HTML}{000080}

\colorlet{blue1}{cyan!30!white}
\colorlet{blue2}{blue}
\colorlet{blue3}{blue!50}
\colorlet{blue4}{blue!20!yellow}
\colorlet{blue5}{yellow!80!orange}
\colorlet{blue6}{red!50!yellow}
\colorlet{blue7}{red}

\colorlet{grayfill}{gray!50}

\pgfplotsset{
    discard if not/.style 2 args={
        model filter/.code={
            \edef\tempa{\thisrow{#1}}
            \edef\tempb{#2}
            \ifx\tempa\tempb
            \else
                
            \fi
        }
    }
}

\newenvironment{customlegend}[1][]{%
    \begingroup
    \csname pgfplots@init@cleared@structures\endcsname
    \pgfplotsset{#1}%
}{%
    \csname pgfplots@createlegend\endcsname
    \endgroup
}%

\def\addlegendimage{\csname pgfplots@addlegendimage\endcsname}

\tikzstyle{globaloptions} = [] 
\tikzstyle{cl} = [draw=blue, mark=*, mark options={scale=0.6, fill=blue, fill opacity=0.7}, globaloptions]  
\tikzstyle{bter} = [draw=red, mark=triangle*, mark options={scale=0.9, fill=red, fill opacity=0.7}, globaloptions]  
\tikzstyle{cnrg} = [draw=green, mark=square*, mark options={scale=0.6, fill=green, fill opacity=0.7}, globaloptions]   
\tikzstyle{sbm} = [draw=brown!50!black, mark=diamond*, mark options={scale=0.8, fill=brown!50!black, fill opacity=0.7}, globaloptions]    
\tikzstyle{er} = [black, thick, globaloptions] 
\tikzstyle{hrg} = [draw=orange!60!red, mark=star, mark options={scale=1.0, fill=cyan!70!blue, fill opacity=0.9}, globaloptions]  
\tikzstyle{kron} = [draw=cyan!50!magenta, mark=x, mark options={scale=1.1, fill=cyan!50!magenta, fill opacity=0.9}, globaloptions]  

\tikzstyle{bugge} = [draw=blue, dashed, mark=*, mark options={solid, fill=white, scale=0.6}, globaloptions]  
\tikzstyle{netgan} = [draw=green, dashed, mark=square, mark options={solid, scale=0.6}, globaloptions]  
\tikzstyle{linae} = [draw=red, dashed, mark=triangle*, mark options={solid, fill=white, scale=0.9}, globaloptions]   
\tikzstyle{gcnae} = [draw=cyan!70!blue, dashed, mark=+, mark options={solid, fill=white, scale=1.1}, globaloptions] 
\tikzstyle{graphrnn} = [draw=cyan!50!magenta, dashed, mark=diamond, mark options={solid, scale=1.1}, globaloptions]  

\tikzstyle{opacityoptions} = [thin, fill opacity=0.95, draw=black]
\tikzstyle{cliquering}=[mark=*, only marks, opacityoptions]
\tikzstyle{condmat}=[mark=*, only marks, opacityoptions]
\tikzstyle{tree}=[mark=diamond*, only marks, opacityoptions]
\tikzstyle{enron}=[mark=diamond*, only marks, opacityoptions]
\tikzstyle{eucore}=[mark=pentagon*,only marks, opacityoptions]
\tikzstyle{flights}=[mark=triangle*, only marks, opacityoptions]
\tikzstyle{chess}=[mark=square*, only marks, opacityoptions]

\usepackage[eulergreek]{sansmath}
\pgfplotsset{
  tick label style = {font=\sansmath\sffamily},
  every axis label = {font=\sansmath\sffamily},
  legend style = {font=\sansmath\sffamily},
  label style = {font=\sansmath\sffamily},
  title style = {font=\sansmath\sffamily}
}
\tikzset{every picture/.style={/utils/exec={\sffamily}}}

\pgfmathdeclarefunction{lg10}{1}{%
    \pgfmathparse{ln(#1)/ln(10)}%
}
\pgfplotscolorbarCMYKworkaroundfalse

\begin{document}
\bstctlcite{bst-control}
%

\title{The Infinity Mirror Test for Graph Models}

\author{Satyaki~Sikdar,
    Daniel~Gonzalez~Cedre,
    Trenton~W.~Ford,
    and Tim~Weninger
\IEEEcompsocitemizethanks{\IEEEcompsocthanksitem[] S. Sikdar, D. Gonzalez Cedre, T. W. Ford and T. Weninger are with the Department of Computer Science \& Engineering at the University of Notre Dame. Email: \{ssikdar,dgonza26,tford5,tweninge\}@nd.edu.}
\thanks{Manuscript received ; revised .}}

\markboth{
}%
{Sikdar \MakeLowercase{\textit{et al.}}: The Infinity Mirror Test for Graph Models}
%



\IEEEtitleabstractindextext{%
\begin{abstract}
Graph models, like other machine learning models, have implicit and explicit biases built-in, which often impact performance in nontrivial ways. The model's faithfulness is often measured by comparing the newly generated graph against the source graph using any number of graph properties. Therefore, differences in the size or topology of the generated graph indicate a loss in the model. Yet, in many systems, errors encoded in loss functions are subtle and not well understood. In the present work, we introduce the \textit{Infinity Mirror} test for analyzing the robustness of graph models. This straightforward stress test works by repeatedly fitting a model to its outputs. A hypothetically perfect graph model would have no deviation from the source graph; however, a model's implicit biases and assumptions are exaggerated by the Infinity Mirror test, exposing potential previously obscured issues. Through an analysis of thousands of experiments on synthetic and real-world graphs, we show that several conventional graph models degenerate in exciting and informative ways. We believe that the observed degenerative patterns are clues to the future development of better graph models.
\end{abstract}

}

\maketitle

\IEEEdisplaynontitleabstractindextext

%
\IEEEpeerreviewmaketitle

\IEEEraisesectionheading{\section{Introduction} \label{sec:intro}}

\IEEEPARstart{M}{eaningful} information is often hidden in subtle interactions and associations within data.
Naturally, graphs are well-suited to representing the connectivity structures that emerge from many real-world social, biological, and physical phenomena. Often, to gain a deeper understanding of a graph's topology, it is useful to summarize a graph through specific representative characteristics that capture its structure. These summarizations often abstract some graph details and no longer represent any single graph but rather an entire set of graphs sharing similar characteristics.
The faithfulness of a graph model on an input graph is usually tested by asking a model to make predictions about the graph's evolution or by generating a new graph using some production scheme.
In the generative case, if the model faithfully captures the source graph's structure, the subsequently generated graphs should resemble the original according to some similarity criteria. 

These graph models come in many varieties.
For example, the Erd\H{o}s–R\'{e}nyi model relies only on external parameters---typically a count of nodes and edges---to determine how it will randomly connect nodes, making it incapable of truly \textit{learning} any deeper topological structure~\citep{erdos1960evolution}.
More recent graph models, like Chung-Lu's Configuration model, improve on the Erd\H{o}s–R\'{e}nyi model by combining specific extrinsic parameters with information learned directly from the input graph~\citep{chung2002average}.
Then, those models, including grammar-based schemes and graph neural networks, are parameterized solely by the topology of the source graph.
These latter two classes of graph models seek a more comprehensive approach by imbuing their production strategies with salient topological information extracted directly from the source graph.

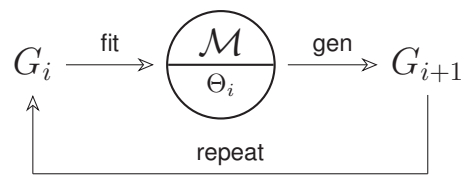
\begin{figure}[t]
    \centering
    \begin{tikzpicture}[tips=proper]

\node [state with output, thick, outer sep=5pt] (param) at (0, 0) {\Large $\mathcal{M}$ \nodepart{lower} $\Theta_i$};
\node [textnode, outer sep=5pt] (gi) at (-2.5,0) {\Large $G_i$};
\node [textnode, outer sep=5pt] (gj) at (2.75,0) {\Large $G_{i+1}$};

\draw [-{Stealth[scale=1.3,angle'=45,open]}] (gi.east) edge node[above=1.75pt, midway, sloped,]{ \small{\textsf{fit}}} (param.west);

\draw [-{Stealth[scale=1.3,angle'=45,open]}] (param.east) edge node[midway, above, sloped] { \small{\textsf{gen}}} (gj.west);

\node [] (westdot) at (-2.5,-1.45) {};
\node [] (eastdot) at (2.75,-1.45) {};

\draw [-{Stealth[scale=1.3,angle'=45,open]}] (westdot.center) edge (gi.south) ;
\draw [] (gj.south) edge (eastdot.center);
\draw [] (westdot.center) edge node[midway, above, sloped] {\small{\textsf{repeat}}} (eastdot.center);

\end{tikzpicture}
    \caption{The Infinity Mirror test iteratively fits a model $\mathcal{M}$ on graph $G_i$, uses the fit parameters $\Theta_i$ to generate a new graph $G_{i + 1}$, and repeats with $G_{i + 1}$. Model biases and errors are quickly revealed.}
    \label{fig:framework}
\end{figure}

Just as statistical biases exist in classical machine learning models, any graph model will make implicit assumptions and value judgments that influence the learning and generation process. 
Of course, this is not necessarily undesirable; principled assumptions are often necessary for decision-making. Certain aspects of graphs may be more important based on the model wielder's focus and intentions.
Indeed, models like Chung-Lu---which learns a degree distribution---and the various Stochastic Block Models, which capture clustering information, are clear about which graph properties they preserve and ignore.

One must then ask: what assumptions do graph neural networks make when learning parameters from a source graph?
What implicit biases drive a grammar-based model to extract one production rule over another?
These questions are often difficult to answer, and traditional methodologies may not readily reveal these hidden inclinations.

This paper presents the Infinity Mirror test: a framework for revealing and evaluating statistical biases present in graph models. The Infinity Mirror test takes its name from the children's toy, which contains two mirrors that reflect light interminably between them.
As illustrated in Fig.~\ref{fig:framework}, this framework operates by iteratively applying a particular graph model onto a graph that it previously generated, constructing a sequence of generated graphs starting from some source graph.
Like how a JPEG image reveals compression artifacts when repeatedly re-compressed, a graph will degenerate when the same model repeatedly fits its outputs.
This sequence of generated graphs can be analyzed using a variety of graph similarity metrics to quantify how the generated graphs diverge from the source graph.
Suppose the sequence is allowed to grow long enough. In that case, this repetition is likely to cause the sequence of graphs to deviate from the source in a way that exposes unknown statistical biases hidden in the model.

\section{Preliminaries} \label{sec:prelim} 
A graph $G = (V,E)$ is defined by a finite set of nodes $V$ and a set of edges $E$.
We denote a node by $v_i \in V$ and an edge between $v_i$ and $v_j$ is given by $e_{ij} = (v_i, v_j) \in E$.
For convenience, let $n = |V|$ and $m = |E|$.
It is sometimes desirable to represent the graph as an $n \times n$ adjacency matrix $A = [a_{ij}]$, where $a_{ij} = 1$ if $e_{ij} \in E$ and $0$ otherwise.
We take the convention that all graphs are undirected, static, and unlabeled unless otherwise indicated.



\subsection{Graph Models} 
A graph model $\mathcal{M}$ is any process or algorithm by which a set of salient features $\Theta$ can be extracted from a graph $G$.
In prediction scenarios, the performance of $\mathcal{M}$ can be assessed using standard precision and recall metrics on held-out data.
If $\mathcal{M}$ also describes how new graphs can be constructed from $\Theta$, its performance can be analyzed by comparing the generated graphs to $G$ using various measures of graph similarity.

Early graph models like the random graph of Erd\H{o}s and R\'{e}nyi~\citep{erdos1960evolution}, the small world network of Watts and Strogatz~\citep{watts1998collective}, the scale-free graph of Albert and Barab\'{a}si~\citep{barabasi1999emergence} and its variants~\citep{bianconi2001competition,ravasz2003hierarchical}, or the more recent LFR benchmark graph generators~\citep{lancichinetti2008benchmark}  generate graphs by applying hand-tuned parameters to some underlying generative process.
This exercise of fine-tuning the model parameters to generate topologically faithful graphs to an input graph is taxing and often hard to achieve. In response, graph models were developed to automatically learn the topological properties of the source graph for a more faithful generation.

One of the first of this new generation of graph models was Chung-Lu's Configuration model~\citep{chung2002average}.
It generated graphs by randomly rewiring edges based on the degree sequence of the source graph.
Even though the degree sequence of the generated graph exactly matched that of the original, the model often failed to incorporate higher-order topological structures like triangles, cliques, cores, and communities observed in the original graph. Since its introduction, more comprehensive models have attempted to fix these flaws by proposing improvements like incorporating assortativity and clustering~\citep{pfeiffer2012fast,mussmann2014assortativity,mussmann2015incorporating,kolda2014scalable}.

For example, the Block Two-level Erd\H{o}s-R\'{e}nyi (BTER) model interprets its input as a scale-free collection of dense Erd\H{o}s-R\'{e}nyi graphs~\citep{kolda2014scalable}.
BTER respects two properties of the original graph: local clustering and degree sequence.
However, BTER sometimes fails to capture higher-order structures and degenerates in graphs with homogenous degree sequences (\eg, grids).
Stochastic Block Models (SBMs) primarily consider communities in the source graph and then create a block matrix that encodes communities as block-to-block connectivity patterns~\citep{karrer2011stochastic, funke2019stochastic}. To generate a graph, the SBM creates an Erd\H{o}s-R\'{e}nyi graph inside each block and random bipartite graphs across communities.
Since SBMs' introduction, they have been extended to handle edge-weighted~\citep{aicher2013adapting}, bipartite~\citep{larremore2014efficiently}, temporal~\citep{peixoto2015inferring}, and hierarchical networks~\citep{peixoto2014hierarchical}.
Likewise, Exponential Random Graph Models (ERGMs)~\citep{robins2007introduction}, Kronecker graph models~\citep{leskovec2010kronecker,leskovec2007scalable,gleich2012moment}, and graph grammar models~\citep{aguinaga2018learning, sikdar2019modeling, hibshman2019towards} are able to generate graphs that are more-or-less faithful to the source graph.

Recent advances in graph neural networks have produced graph generators based on recurrent neural networks~\citep{you2018graphrnn}, variational autoencoders~\citep{salha2020simple, kipf2016variational, li2018learning}, transformers~\citep{yun2019graph}, and generative adversarial networks~\citep{bojchevski2018netgan}, each of which has advantages and disadvantages.
Graph autoencoders (GAE) learn an embedding of the input graph's nodes via message-passing and then construct new graphs by taking inner products of the embedding vectors and passing them through an activation function (\eg, sigmoid).
NetGAN (NGAN) trains a Generative Adversarial Network (GAN) to \textit{generate} and \textit{discriminate} between real and synthetic random walks over the input graph. After training, the model builds graphs from random walks produced by the generator.
GraphRNN (GRNN), a kind of Recurrent Neural Network (RNN), decomposes the process of graph generation into two separate RNNs---one for generating a sequence of nodes and the other for the sequence of edges. 

\begin{figure}[tb]
    \centering
    \includegraphics[width=0.99\linewidth]{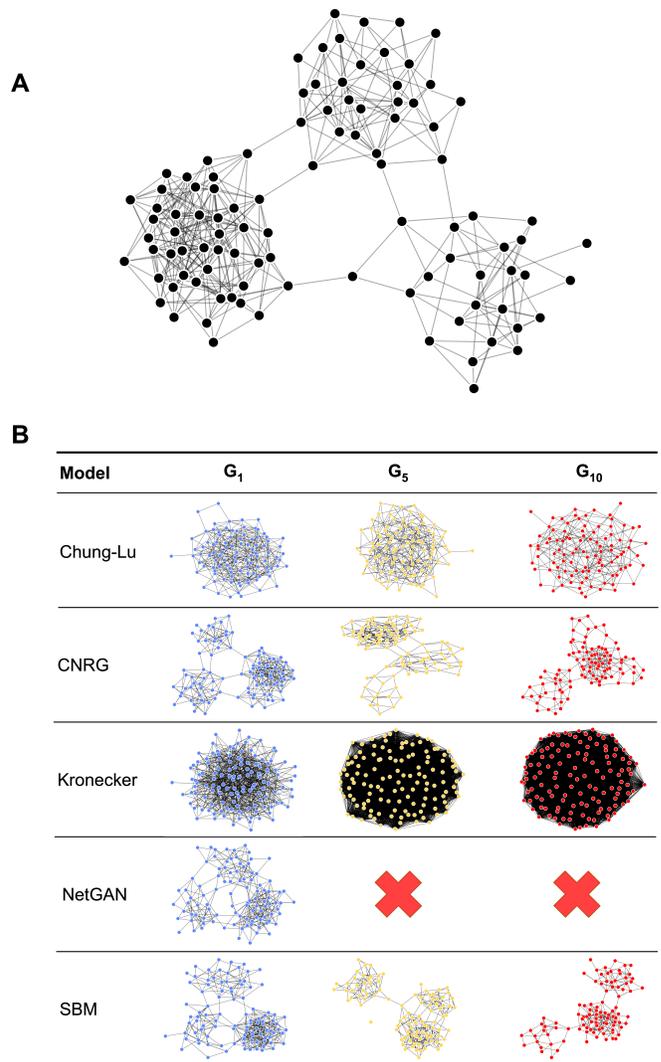}
    \vspace{.3cm}
    \caption{(A) Example graph with three distinct communities. (B) Graphs  $G_1$, $G_5$, and $G_{10}$ are generated using the Infinity Mirror framework described in Fig.~\ref{fig:framework} for various models. Chung-Lu and Kronecker immediately lose the community structure of the source graph. Kronecker progressively makes the graph denser. CNRG and SBM retain the input graph's community structure, albeit with some appreciable deterioration.}
    \label{fig:toy-example}
    \vspace{-.2cm}
\end{figure}

\subsection{Graph Comparison Metrics} 
Graph models are typically evaluated by their ability to predict nodes and edges. Although the prediction task is important, it does not measure a model's ability to capture a graph's topological structure.

Another evaluation approach involves comparing a source graph $G$ to a new graph $\hat{G}$ generated from $\mathcal{M}$ using a measure of graph similarity or divergence. Here, the choice of metric influences the differences and biases that are exposed within the model.

The simplest graph comparison metrics are computed by comparing distributions of first-order graph properties like the graph's degree and PageRank distributions.
In addition to visually inspecting these distributions, graph modelers also employ quantitative metrics like the Jensen-Shannon (JS) divergence.
More advanced metrics compare two graphs by examining properties of their adjacency matrices. There are two conventional approaches in this space: (1) known node-correspondence metrics, which assume every node in $G$ has a known corresponding node in $\hat{G}$, and (2) unknown node-correspondence metrics, where there is no assumed node correspondence. 

\textsc{DeltaCON} is an example of a known node-correspondence metric that compares node affinities between $G$ and $\hat{G}$~\citep{koutra2013deltacon,koutra2016deltacon}. Affinities are measured using belief propagation, which associates every node with a vector measuring its influence on its $k$-hop neighbors. A score is produced by comparing the vectors of corresponding node affinities. 
Unfortunately, most graph models do not generate graphs with a known node-correspondence. Although best-guess node correspondence can be predicted if needed, known node-correspondence metrics like \textsc{DeltaCON} and the cut distance~\citep{liu2018cut} are not well-suited for the present work.

Fortunately, there also exist many unknown node-correspondence metrics. Examples include Portrait divergence and $\lambda$-distance~\citep{bagrow2019information,wilson2008study}.
We can also directly compare the graphlet counts~\citep{ahmed2015efficient} between $G$ and $\hat{G}$, or compute their graphlet correlation distance (GCD)~\citep{prvzulj2007biological} by counting node orbitals.
Finally, loss functions from recent graph neural network models compare $G$ to $\hat{G}$ by collating derived graph features like power-law exponents, diameters, and node centrality scores~\citep{bojchevski2018netgan}.

In each case, the features extracted by a model and the generative process by which predictions are made carry inherent biases, which may elude even the most comprehensive performance or comparison metric.

\section{Infinity Mirror test} \label{sec:method}

The Infinity Mirror~\citep{aguinaga2016infinity} test seeks to expose a graph model's implicit biases by iteratively computing hereditary chains of graphs generated by the model. For a graph model $\mathcal{M}$ and source graph $G_0$, we define a chain $\langle G_1, G_2, \dots, G_\ell \rangle$ of length $\ell$ by computing: 
$$G_{i + 1} = \mathcal{M}(G_i, \Theta_i)$$ 
\noindent at each iteration $i$ of the chain, where $\Theta_i$ denotes the features extracted from $G_i$ by the model $\mathcal{M}$. Fig.~\ref{fig:framework} illustrates this iterative fit-and-generate process. 

Because each subsequent graph is a lossy reflection of a previous graph, which itself may have been a lossy reflection of one prior, we expect that this chain of graphs will diverge from the source graph as the chain grows longer. See Fig.~\ref{fig:toy-example} for an example. By inspecting the divergence along the chain of generated graphs, patterns of model error or bias should become easier to detect.

This chain of graphs can be evaluated from several different perspectives. The most straightforward way to measure the error in a chain is by comparing the initial graph $G_0$ to the last graph $G_\ell$ using a graph similarity metric.
This provides insight into the total degradation resulting from model-specific errors and biases.




Analogously, analyzing the features $\Theta_i$ that a model learns when generating a chain might shine a light on the inner workings of a model, which might not be reflected by merely comparing output graphs to each other. 
However, this would require a more tailored approach than we can provide in the present work as different generative models learn different sets of features that are often incomparable. 

\para{Kronecker Graphs.} Next, we apply the Infinity Mirror test on the Kronecker graph model, which is known to produce graphs that approximate a log-normal degree distribution (with a heavy tail)~\citep{seshadhri2013depth}. Although this property of the Kronecker graph model is often desirable, it is no doubt a bias that is encoded into the model. If we provide a source graph that does not follow a log-normal degree sequence, we expect the degree sequence to diverge relative to the source graph. 

\begin{figure}
    \centering
        \begin{tikzpicture}
\begin{scope}[scale=0.6, transform shape, shift={(-3.8752,0)}]
\begin{scope}[shift={(1.5827,0)}]
	\node[inode, ellipse, minimum height=50pt, minimum width=100pt, outer sep=5pt] (a) at (0,0) {};
	\node[inode, ellipse, minimum height=50pt, minimum width=100pt, outer sep=5pt] (d) at (5,0) {};
	
	\draw [diredge] (a) edge [loop above, looseness=5] node {\large $a$} (a);
	\draw [diredge] (a) edge [bend left] node [above] {\large $b$} (d);
	\draw [diredge] (d) edge [bend left] node [above] {\large $c$} (a);
	\draw [diredge] (d) edge [loop above, looseness=5] node {\large $d$} (d);
\end{scope}

\begin{scope}[scale=0.8, transform shape, shift={(0.8261,-0.2376)}]
	\node[inode, fill=gray!30!white, ellipse, minimum height=15pt, minimum width=30pt, outer sep=2pt] (a) at (0,0) {$\cdots$};
	\node[inode, fill=gray!30!white, ellipse, minimum height=15pt, minimum width=30pt, outer sep=2pt] (d) at (2.2084,0) {$\cdots$};
	
	\draw [diredge] (a) edge [loop above, looseness=5] node {\large $a \cdot a$} (a);
	\draw [diredge] (a) edge [bend left] node [above, looseness=6] {\large $a \cdot b$} (d);
	\draw [diredge] (d) edge [bend left] node [above] {\large $a \cdot c$} (a);
	\draw [diredge] (d) edge [loop above, looseness=6] node {\large $a \cdot d$} (d);
\end{scope}

\begin{scope}[scale=0.8, transform shape, shift={(7.1631,-0.2376)}]
	\node[inode, fill=gray!30!white, ellipse, minimum height=15pt, minimum width=30pt, outer sep=2pt] (a) at (0,0) {$\cdots$};
	\node[inode, fill=gray!30!white, ellipse, minimum height=15pt, minimum width=30pt, outer sep=2pt] (d) at (2.2916,0) {$\cdots$};
	
	\draw [diredge] (a) edge [loop above, looseness=5] node {\large $d \cdot a$} (a);
	\draw [diredge] (a) edge [bend left] node [above] {\large $d \cdot b$} (d);
	\draw [diredge] (d) edge [bend left] node [above] {\large $d \cdot c$} (a);
	\draw [diredge] (d) edge [loop above, looseness=5] node {\large $d \cdot d$} (d);
\end{scope}

\begin{scope}[shift={(0,0.4998)}]
\node [textnode] at (-3.7501,0) {\Large $\mathcal{I}_{2 \times 2} = \begin{bmatrix} a & b \\ c & d \end{bmatrix}$};
\node at (-3.4167,-1.7499) {\Large $A_{2^n \times 2^n} = \underbrace{\mathcal{I} \otimes \cdots \otimes \mathcal{I}}_{n \text{ times}} $};

\end{scope}
\end{scope}

\node [textnode] at (-5.4,1) {\Large\textbf{\textsf{A}}};
\node [textnode] at (-2.5,1) {\Large\textbf{\textsf{B}}};

\end{tikzpicture}
        \vspace*{0.5em}
        \input{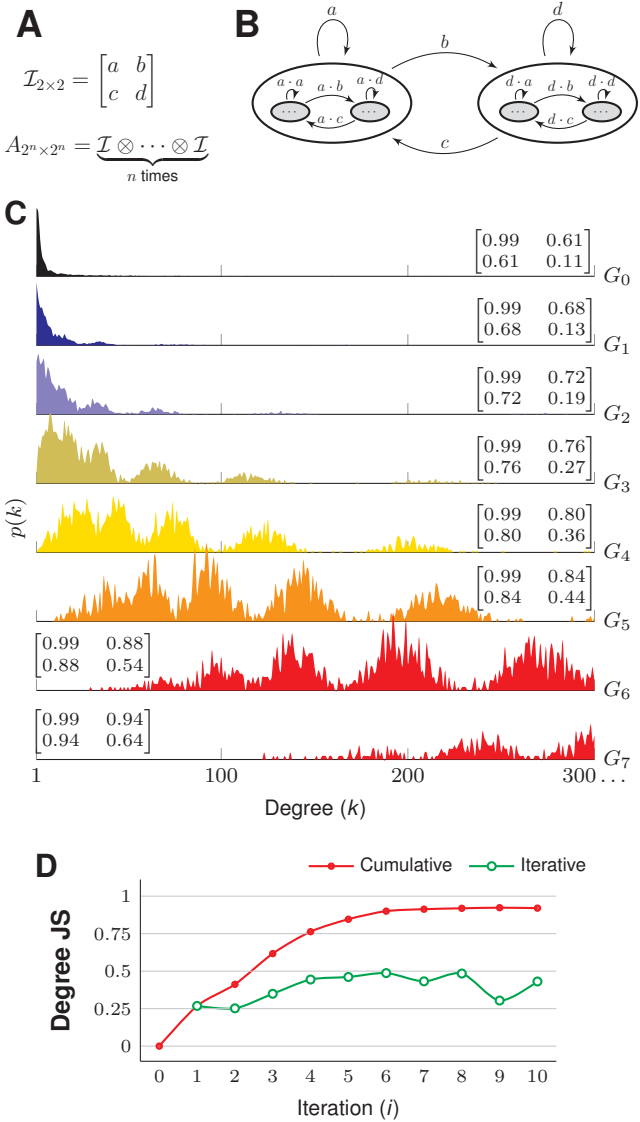}
        \vspace*{0.75em}
        \pgfplotstableread{
gen abs seq
1   0.078   0.078
2   0.111   0.112
3   0.198   0.103
4   0.253   0.055
5   0.251   0.013
6   0.133   0.17
7   0.102   0.063
8   0.101   0.016
9   0.102   0.017
10  0.1 0.013
11  0.104   0.021
12  0.101   0.027
13  0.107   0.015
14  0.103   0.022
15  0.102   0.039
16  0.099   0.031
17  0.102   0.015
18  0.101   0.023
19  0.103   0.009
20  0.104   0.023
}{\cvm} 

\pgfplotstableread{
gen abs seq
1   0.431   0.431
2   1.017   0.426
3   2.757   0.826
4   5.262   1.331
5   7.382   1.468
6   9.027   1.697
7   9.161   1.355
8   9.382   1.553
9   9.497   0.589
10  9.598   1.352
11  9.598   0.562
12  9.453   0.481
13  9.694   0.595
14  9.701   0.386
15  9.671   0.44
16  9.709   0.677
17  9.76    0.413
18  9.692   0.474
19  9.555   0.419
20  9.684   0.364
}{\flightskl}

\pgfplotstableread{
gen abs seq
0   0   nan
1   0.268   0.268
2   0.412   0.252
3   0.617   0.349
4   0.763   0.444
5   0.846   0.461
6   0.9 0.487
7   0.913   0.432
8   0.919   0.484
9   0.923   0.304
10  0.92    0.431
11  0.921   0.298
12  0.916   0.283
13  0.925   0.312
14  0.924   0.253
15  0.922   0.273
16  0.925   0.332
17  0.927   0.263
18  0.923   0.279
19  0.92    0.264
20  0.924   0.249
}{\flightsjs}

\begin{tikzpicture}

\begin{scope}[shift={(-2, -1)}]
\begin{axis}[
    height=4.0cm,
    width=0.8\linewidth,
    tick label style={font=\scriptsize}, 
    every axis title/.style={below left,at={(1,1),anchor=west},font=\footnotesize},
    ylabel style={align=center},
    ylabel={\textbf{Degree JS}},
    xlabel={\footnotesize{\textsf{Iteration ($i$)}}},
    yticklabel style={
            /pgf/number format/fixed,
            /pgf/number format/precision=5
    },
    scaled y ticks=false,
    ymin=0,
    ymax=1.01,
    xmin=0,
    xmax=10,
    restrict x to domain=0:10,
    enlarge x limits=0.05, 
    enlarge y limits=0.1,
    ylabel style={align=center, yshift=0.05em},
    ymajorgrids,
    axis lines*=left,
    x tick style=transparent,
    y tick style=transparent,
    axis on top,
    legend cell align={left},
    legend columns=2,
    legend style={at={(axis cs:3.7,1.2)},anchor=west, font=\scriptsize, draw=none, align=left, fill=none},
    xtick={0,1,2,3,4,5,6,7,8,9,10},
    ytick={0,0.25, 0.5, 0.75, 1},
    ]

    \addplot [draw=red, mark options={scale=1.0, fill=red, fill opacity=0.7}, mark=*, mark size=1pt, thick, smooth]  table [x=gen, y=abs]   {\flightsjs};
    \addplot [draw=green, smooth, mark=*, mark size=1.5pt, mark options={fill=white}, thick]  table [x=gen, y=seq]   {\flightsjs};
    
    \legend{\textsf{Cumulative~~}, \textsf{Iterative~~}}
\end{axis}
\end{scope}

\node [textnode] at (-3.25,1.55) {\large\textbf{\textsf{D}}};
\end{tikzpicture}
    \caption{
    (A) A generic $2\times 2$ Kronecker initiator matrix $\mathcal{I} = [a\, b; c\, d]$ with $0 \le a,b,c,d < 1$ can be used to generate an adjacency matrix of size $2^n \times 2^n$ by using Kronecker products. (B) State transition diagram corresponding to the initiator matrix $\mathcal{I}$ in (A) visually representing the recursive growth process of Kronecker graphs. (C) Degree distribution on the Flights graph illustrated over seven fit-and-generate iterations of the Kronecker model. The Kronecker initiator matrix from \texttt{KronFit} is positioned over the individual ridge plots. (D) Jensen-Shannon divergence of the $i^\textrm{th}$ graph compared to the source graph $JS(G_0,G_i)$ (red line) and of the $i^\textrm{th}$ graph compared to the $i-1^\textrm{th}$ graph $JS(G_{i-1},G_i)$ (green line). This example demonstrates a hidden bias in the Kronecker graph model: the degree distribution tends to flatten and oscillate after a few iterations. 
    }
    \label{fig:kronexample}
\end{figure}

In the Kronecker model, graphs are modeled as a repeated Kronecker product of a $k \times k$ initiator matrix $\mathcal{I}$ (usually, $k$ = $2$) as shown in Fig.~\ref{fig:kronexample}(A). Every entry of $\mathcal{I}$ can be thought of as a transition probability between two states, as seen in Fig.~\ref{fig:kronexample}(B). The \texttt{KronFit}\footnote{\texttt{KronFit} and \texttt{KronGen} can be obtained from SNAP} utility can be used to fit the initiator matrix to an input graph, while the \texttt{KronGen} utility can be used to generate new graphs from an initiator matrix, by performing repeated Kronecker products of the initiator matrix. As a result, it can only generate graphs with nodes which are in powers of $k$.

For example, consider the plots in Fig.~\ref{fig:kronexample}(C) which illustrate the degree distributions of a chain of graphs obtained by performing the Infinity Mirror test on a graph of airline flights\footnote{\url{https://openflights.org/data.html}} using the Kronecker model. 
The subplot labeled $G_0$ shows the degree distribution of the original Flights graph. We then fed the Flights graph to \texttt{KronFit} and generated a graph from the learned initiator matrix (\texttt{KronGen}) to create a new graph $G_1$~\citep{leskovec2010kronecker}. The plot labeled $G_1$ illustrates the degree distribution of this new graph. The degree distributions of $G_0$ and $G_1$ look visually similar. The next step is to compute some distribution similarity measure to analytically compare the two distributions (we use Jensen-Shannon (JS) divergence), concluding that the graph model has some error $\epsilon$.

Rather than stopping here, the Infinity Mirror methodology continues this fit-and-generate process. So, we input $G_1$ into the Kronecker model and generated a new graph $G_2$, whose degree distribution is illustrated in the plot labeled $G_2$. Likewise, the plots labeled $G_3,G_4,\ldots, G_7$ illustrate the degree distribution of subsequent iterations. A visual inspection of these plots shows that the degree sequence of the Kronecker model degenerates into an ever-spreading oscillating distribution. We also note that the entries in the initiator matrices monotonically increase in magnitude across generations, signifying the increase in edge density~\citep{leskovec2010kronecker}.

Fig.~\ref{fig:kronexample}(D) provides an analytical view of the visual plots on top as the number of iterations continues to $10$. The cumulative JS divergence compares the degree distribution of $G_0$ to $G_i$, thereby accounting for the error over multiple iterations. Conversely, the iterative JS divergence compares the degree sequence of $G_{i-1}$ to $G_{i}$, accounting for errors in each iteration. The iterative error shows that each fit and generate procedure produces some amount of error, which is expected when performing any modeling. However, it is important to note that the cumulative error is not simply a summation of each iteration error. It is certainly possible for an iterative error made by some later iteration to correct an earlier error and lead to a decrease in the cumulative error. In this case, we find that the cumulative error starts off low, but diverges asymptotically. 

In summary, this iterative test can not only capture model-induced error, it can also reveal bias encoded in the model, such as the Kronecker model's dispersing degree distribution previously uncovered and formalized by Seshadhri \etal~\citep{seshadhri2013depth}.

\begin{figure*}
    \centering
    \input{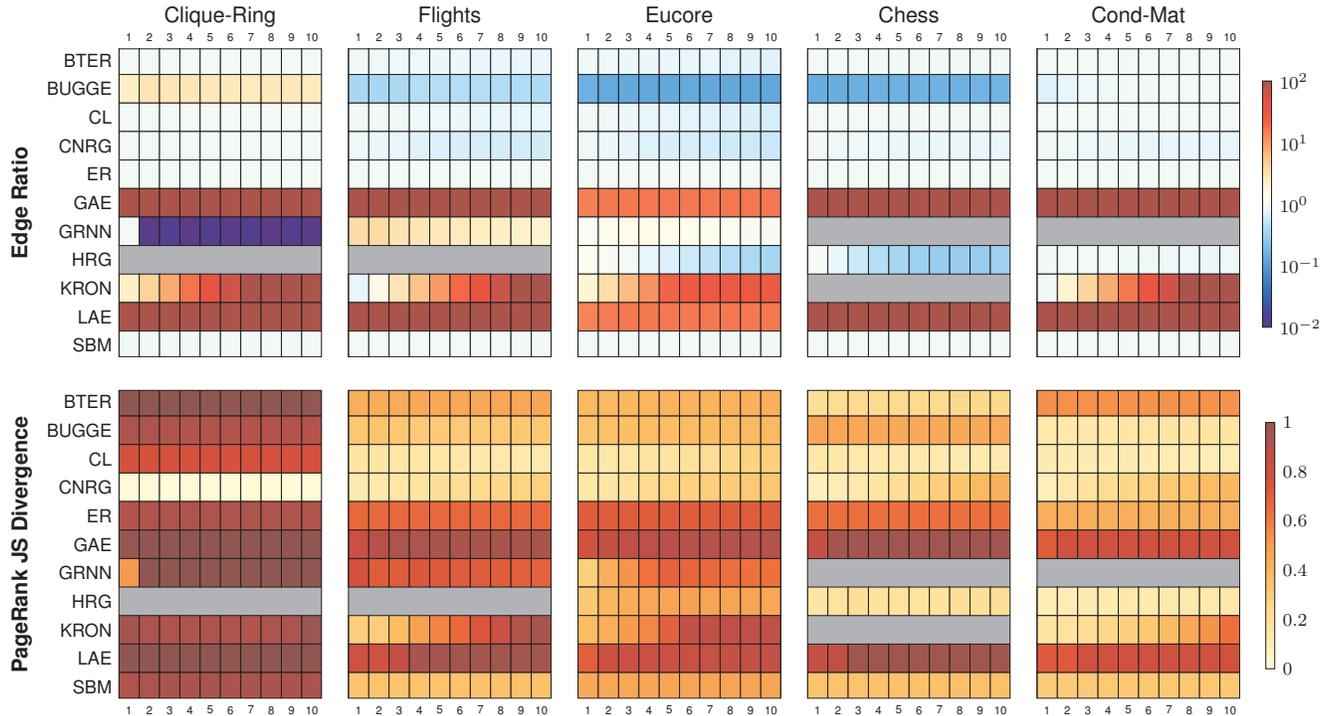}
\pgfplotsset{colormap/temp}
\pgfplotsset{colormap/YlOrRd}
\pgfplotsset{colormap/temp}

\begin{tikzpicture}[scale=0.85, transform shape]
    \begin{groupplot}[
        group style={
            group size=5 by 3,
            horizontal sep=12pt,
            vertical sep=15pt,
            x descriptions at=edge bottom,
            y descriptions at=edge left,
        },
        width=4.75cm,
        height=6.4cm,
        enlarge x limits={rel=0.05, upper},
        enlarge y limits=0.04,      
        nodes near coords style={
            scale=0.75,
            yshift=-0.6em,
            text=white,
            /pgf/number format/.cd,
            fixed, fixed zerofill,
            precision=2,
        },
        colormap name=YlOrRd,
        ytick=data,
        xtick=data,
        ytick style={draw=none},
        yticklabel style = {font=\footnotesize,xshift=0.05ex},
        xticklabel style = {font=\tiny},
        xtick style={draw=none},
        xlabel=Iteration,
        title style={yshift=0.25em},
        axis background/.style={fill=gray!70}
    ]
    
    \nextgroupplot[
        xticklabel pos=top,
        title=Clique-Ring,
        title style={yshift=-0.19em},
        colormap name=temp,
        point meta min=-2,
        point meta max=2,
        ylabel={\textbf{Edge Ratio}},
        symbolic x coords={1,2,3,4,5,6,7,8,9,10},
        symbolic y coords={BTER,BUGGE,CL,CNRG,ER,GAE,GRNN,HRG,KRON,LAE,SBM},
    ]
    \addplot[
        matrix plot,
        draw=black,
        fill opacity=1,
        mark=none,
        mesh/cols=10,
        point meta=explicit,
    ] 
    table[x=iter, y=model, meta expr=log10(\thisrow{frac-m})] {\cliqueringDegPRHeatMap};   
    
    
    \nextgroupplot[
        xticklabel pos=top,
        title=Flights,
        title style={yshift=-0.19em},
        colormap name=temp,
        point meta min=-2,
        point meta max=2,
        symbolic x coords={1,2,3,4,5,6,7,8,9,10},
        symbolic y coords={BTER,BUGGE,CL,CNRG,ER,GAE,GRNN,HRG,KRON,LAE,SBM},
        yticklabels={},
    ]
    \addplot[
        matrix plot,
        draw=black,
        fill opacity=1,
        mark=none,
        mesh/cols=10,
        point meta=explicit,
    ] 
    table[x=iter, y=model, meta expr=log10(\thisrow{frac-m})] {\flightsDegPRHeatMap};   
    
    
    \nextgroupplot[
        xticklabel pos=top,
        title=Eucore,
        colormap name=temp,
        point meta min=-2,
        point meta max=2,
        symbolic x coords={1,2,3,4,5,6,7,8,9,10},
        symbolic y coords={BTER,BUGGE,CL,CNRG,ER,GAE,GRNN,HRG,KRON,LAE,SBM},
        yticklabels={},
    ]
    \addplot[
        matrix plot,
        draw=black,
        fill opacity=1,
        mark=none,
        mesh/cols=10,
        point meta=explicit,
    ] 
    table[x=iter, y=model, meta expr=log10(\thisrow{frac-m})] {\eucoreDegPRHeatMap};
    
    \nextgroupplot[
        xticklabel pos=top,
        title=Chess,
        colormap name=temp,
        point meta min=-2,
        point meta max=2,
        symbolic x coords={1,2,3,4,5,6,7,8,9,10},
        symbolic y coords={BTER,BUGGE,CL,CNRG,ER,GAE,GRNN,HRG,KRON,LAE,SBM},
        yticklabels={},
    ]
    \addplot[
        matrix plot,
        draw=black,
        fill opacity=1,
        mark=none,
        mesh/cols=10,
        point meta=explicit,
    ] 
    table[x=iter, y=model, meta expr=log10(\thisrow{frac-m})] {\chessDegPRHeatMap};
    
    \nextgroupplot[
        xticklabel pos=top,
        title=Cond-Mat,
        colormap name=temp,
        point meta min=-2,
        point meta max=2,
        colorbar,
        colorbar style={width=5pt, ytick={-2,-1,0,1,2}, yticklabels={$10^{-2}$,$10^{-1}$,$10^0$,$10^1$,$10^2$},
        scale=0.8, transform shape, yshift=-1.5em},
        symbolic x coords={1,2,3,4,5,6,7,8,9,10},
        symbolic y coords={BTER,BUGGE,CL,CNRG,ER,GAE,GRNN,HRG,KRON,LAE,SBM},
        yticklabels={},
    ]
    \addplot[
        matrix plot,
        draw=black,
        fill opacity=1,
        mark=none,
        mesh/cols=10,
        point meta=explicit,
    ] 
    table[x=iter, y=model, meta expr=log10(\thisrow{frac-m})] {\condmatDegPRHeatMap};
    
    \nextgroupplot[
        ylabel={\textbf{PageRank JS Divergence}},
        point meta min=0,
        point meta max=1,
        symbolic x coords={1,2,3,4,5,6,7,8,9,10},
        symbolic y coords={BTER,BUGGE,CL,CNRG,ER,GAE,GRNN,HRG,KRON,LAE,SBM},
    ]
    \addplot[
        matrix plot,
        point meta min=0,
        point meta max=1,
        draw=black,
        fill opacity=1,
        mark=none,
        mesh/cols=10,
        point meta=explicit,
    ] 
    table[x=iter, y=model, meta=pr-js] {\cliqueringDegPRHeatMap};  
    
    
    \nextgroupplot[
        point meta min=0,
        point meta max=1,
        symbolic x coords={1,2,3,4,5,6,7,8,9,10},
        symbolic y coords={BTER,BUGGE,CL,CNRG,ER,GAE,GRNN,HRG,KRON,LAE,SBM},
        yticklabels={},
    ]
    \addplot[
        matrix plot,
        draw=black,
        fill opacity=1,
        mark=none,
        mesh/cols=10,
        point meta=explicit,
    ] 
    table[x=iter, y=model, meta=pr-js] {\flightsDegPRHeatMap};

    
    \nextgroupplot[
        point meta min=0,
        point meta max=1,
        symbolic x coords={1,2,3,4,5,6,7,8,9,10},
        symbolic y coords={BTER,BUGGE,CL,CNRG,ER,GAE,GRNN,HRG,KRON,LAE,SBM},
        yticklabels={},
    ]
    \addplot[
        matrix plot,
        draw=black,
        fill opacity=1,
        mark=none,
        mesh/cols=10,
        point meta=explicit,
    ] 
    table[x=iter, y=model, meta=pr-js] {\eucoreDegPRHeatMap};
    
    \nextgroupplot[
        point meta min=0,
        point meta max=1,
        symbolic x coords={1,2,3,4,5,6,7,8,9,10},
        symbolic y coords={BTER,BUGGE,CL,CNRG,ER,GAE,GRNN,HRG,KRON,LAE,SBM},
        yticklabels={},
    ]
    \addplot[
        matrix plot,
        draw=black,
        fill opacity=1,
        mark=none,
        mesh/cols=10,
        point meta=explicit,
    ] 
    table[x=iter, y=model, meta=pr-js] {\chessDegPRHeatMap};
    
    \nextgroupplot[
        point meta min=0,
        point meta max=1,
        colorbar,
        colorbar style={width=5pt, xshift=0.15em, yshift=-1.5em, scale=0.8, transform shape,},
        symbolic x coords={1,2,3,4,5,6,7,8,9,10},
        symbolic y coords={BTER,BUGGE,CL,CNRG,ER,GAE,GRNN,HRG,KRON,LAE,SBM},
        yticklabels={},
    ]
    \addplot[
        matrix plot,
        draw=black,
        fill opacity=1,
        mark=none,
        mesh/cols=10,
        point meta=explicit,
    ] 
    table[x=iter, y=model, meta=pr-js] {\condmatDegPRHeatMap};
    
    \end{groupplot}
\end{tikzpicture}
    \caption{The Edge Ratio (top), and Jensen-Shannon (JS) divergence of the PageRank (bottom) distributions over 10 iterations. Each square in the heatmaps represents the value averaged over 50 trials. An edge ratio of 1 indicates no change. For JS divergence, lower is better. Model failures are marked in gray. Most models preserve the edge counts of the original graphs, except KRON, GRNN, LAE, and GAE. BTER, CL, CNRG, and SBM are standout performers, as evidenced by their low divergence values across the board.}
    \label{fig:degree}
\end{figure*}

\section{Experiments}
In this section, we apply the Infinity Mirror test to several graph models using relevant graph properties and metrics.
The Infinity Mirror test reveals interesting error patterns in many graph models and also suggests hidden biases worthy of further investigation.
Our open-source implementation can be found on GitHub\footnote{\url{www.github.com/satyakisikdar/infinity-mirror}}.

\para{Data.} We perform experiments over five source graphs, denoted $G_0$. Four of these are real-world graphs commonly found in the graph modeling literature: OpenFlights (Flights), an email graph from a large European research institution (Eucore), a network of chess competitions (Chess), a collaboration network of Condensed Matter physics (Cond-Mat). The fifth, Clique-Ring, is a synthetic graph consisting of a 500 node ring where each node is replaced by a 4 clique. Summary statistics on the datasets can be found in Tab.~\ref{tab:dataset}.

\begin{table}[htb]
    \centering
    \caption{Dataset summary statistics: Avg. CC is the average  clustering coefficient, and Avg. PL is the average unweighted shortest path length.}
    \begin{tabular}{@{} l rrr rr @{}}
        \toprule
        \textbf{Dataset} & \textbf{$\mathbf{|V|}$} & \textbf{$\mathbf{|E|}$} & \textbf{\# Triangles} & \textbf{Avg. CC} & \textbf{Avg. PL} \\
        \midrule
        Clique-Ring & 2,000 & 3,500 & 2,000 & 0.75 & 250 \\
        Flights & 2,905 & 15,645 & 72,843 & 0.45 & 4 \\
        Eucore & 986 & 16,064 & 105,461 & 0.40 & 2.5 \\
        Chess & 7,115 & 55,779 & 108,584 & 0.18 & 4 \\
        Cond-Mat & 21,363 & 91,286 & 171,051 & 0.64 & 5.35 \\
        \bottomrule
    \end{tabular}
    \label{tab:dataset}
\end{table}

We also attempted to perform experiments on the Enron email dataset, which contains about 33K notes and 180K edges. Unfortunately, many of the graph models degenerated too quickly (as shown in Dai et al.~\cite{dai2020scalable}) or were otherwise unable to process such a large network.

\para{Graph Models.} There are hundreds of possible graph models to which the Infinity Mirror test may be applied; however, the current work is not a survey of graph models; instead, we sample archetypal graph models from among the various options available. These include Chung Lu (CL)~\citep{chung2002average}, clustering-based node replacement graph grammars (CNRG)~\citep{sikdar2019modeling}, block two-level Erd\H{o}s R\'{e}yni (BTER)~\citep{seshadhri2012bter}, degree-corrected stochastic block models (SBM)~\citep{karrer2011sbm}, hyperedge replacement graph grammars (HRG)~\citep{aguinaga2016growing, aguinaga2018learning}, Kronecker graphs (KRON)~\citep{leskovec2010kronecker, leskovec2007scalable}, bottom-up graph grammar extractor (BUGGE)~\citep{hibshman2019towards}, generative adversarial network (NGAN)~\citep{bojchevski2018netgan}, graph linear autoencoder (LAE)~\citep{salha2020simple}, graph convolutional neural networks (GAE)~\citep{kipf2016variational}, and graph recurrent neural networks (GRNN)~\citep{you2018graphrnn}. 

\para{Methodology.}
For each combination of $\mathcal{M}$ and $G_0$, we create 50 independent fit-and-generate chains, each of length 10 (\ie, $G_1, G_2,\ldots,G_{10}$).
The 50 independent chains are almost certainly different because each graph model incorporates some stochasticity during feature extraction, graph generation, or both.
GraphRNN does not conform to the above format, because it learns from and generates batches of graphs at a time.
For this model, we initially feed in 50 identical copies of $G_0$, which are used to generate a batch of 50 different $G_1$s; in this manner, every iteration of GraphRNN works on batches of 50 graphs for both input and output.

Some graph models like NetGAN degenerate to an extent that they fail to complete even one iteration of the fit-generate process. A thorough investigation of their implementation is needed to unveil the specific causes, which we consider to be beyond the scope of our work. We, therefore, choose to leave out NetGAN from all our experiments. 
In a small number of cases, some models were unable to process some of the graphs (for various reasons); in those cases, results are omitted.

\begin{figure*}
    \centering
    \input{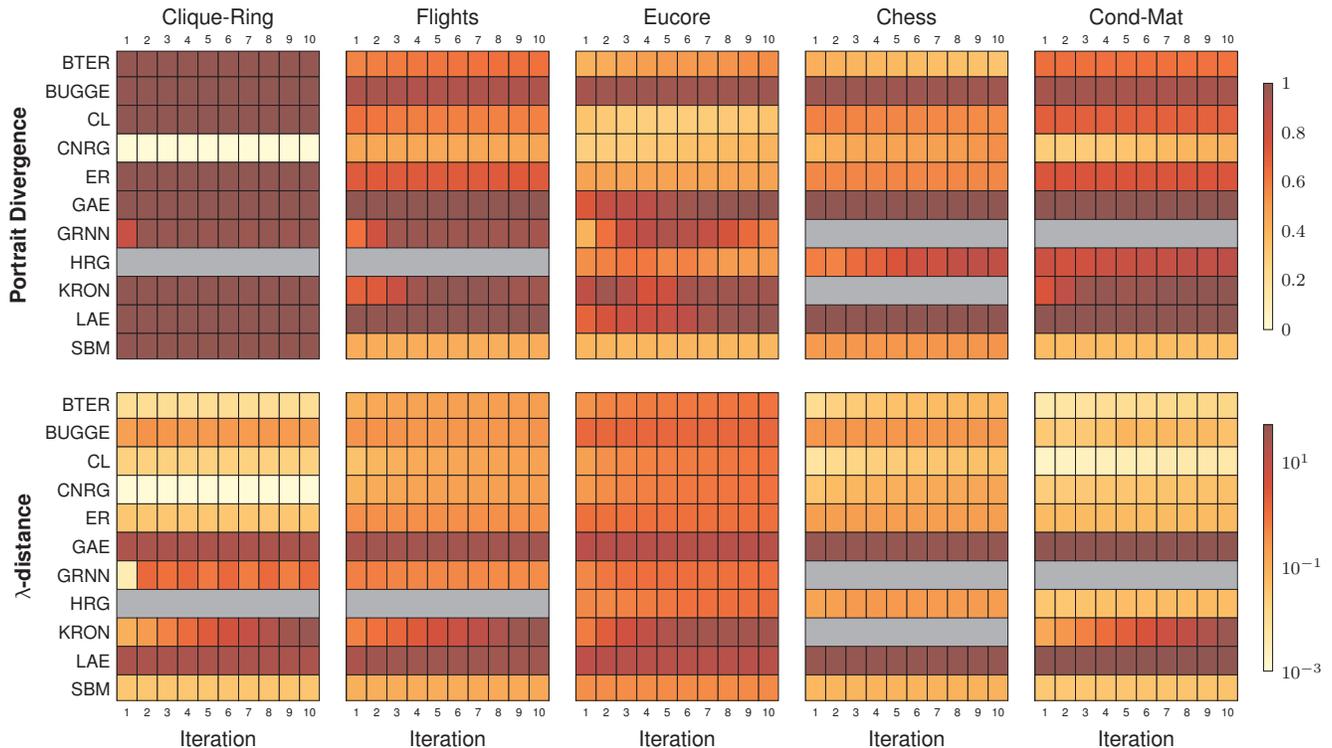}

\pgfplotsset{colormap/YlOrRd}
\pgfplotsset{colormap/jet}

\begin{tikzpicture}[scale=0.85, transform shape]
	\begin{groupplot}[
		group style={
			group size=5 by 2,
			horizontal sep=12pt,
			vertical sep=15pt,
			x descriptions at=edge bottom,
			y descriptions at=edge left,
		},
		width=4.75cm,
		height=6.4cm,
		enlarge x limits={rel=0.05, upper},
		enlarge y limits=0.04,		
		nodes near coords style={
			scale=0.75,
			yshift=-0.6em,
			text=white,
			/pgf/number format/.cd,
			fixed, fixed zerofill,
			precision=2,
		},
		colormap name=YlOrRd,
		point meta min=0,
		point meta max=1,
		ytick=data,
		xtick=data,
		ytick style={draw=none},
		yticklabel style = {font=\footnotesize,xshift=0.1ex},
		xticklabel style = {font=\tiny},
		xtick style={draw=none},
		xlabel=Iteration,
		title style={yshift=0.25em},
		axis background/.style={fill=gray!70}
	]
	\nextgroupplot[
		title=Clique-Ring,
		title style={yshift=-0.175em},
		point meta min=0,
		point meta max=1,
		ylabel={\textbf{Portrait Divergence}},
		symbolic x coords={1,2,3,4,5,6,7,8,9,10},
		symbolic y coords={BTER,BUGGE,CL,CNRG,ER,GAE,GRNN,HRG,KRON,LAE,SBM},
		xticklabel pos=top,
	]
	\addplot[
		matrix plot,
		draw=black, 
		fill opacity=1,
		colorbar source,
		mark=none,
		mesh/cols=10,
		point meta=explicit,
	] 
	table[x=iter, y=model, meta=portrait] {\cliqueringDegPRHeatMap};	
	
	
	\nextgroupplot[
		title=Flights,
		title style={yshift=-0.175em},
		point meta min=0,
		point meta max=1,
		symbolic x coords={1,2,3,4,5,6,7,8,9,10},
		symbolic y coords={BTER,BUGGE,CL,CNRG,ER,GAE,GRNN,HRG,KRON,LAE,SBM},
		yticklabels={},
		xticklabel pos=top,
	]
	\addplot[
		matrix plot,
		draw=black, 
		fill opacity=1,
		colorbar source,
		mark=none,
		mesh/cols=10,
		point meta=explicit,
	] 
	table[x=iter, y=model, meta=portrait] {\flightsDegPRHeatMap};	
	
	
	\nextgroupplot[
		title=Eucore,
		point meta min=0,
		point meta max=1,
		symbolic x coords={1,2,3,4,5,6,7,8,9,10},
		symbolic y coords={BTER,BUGGE,CL,CNRG,ER,GAE,GRNN,HRG,KRON,LAE,SBM},
		yticklabels={},
		xticklabel pos=top,
	]
	\addplot[
		matrix plot,
		draw=black, 
		fill opacity=1,
		colorbar source,
		mark=none,
		mesh/cols=10,
		point meta=explicit,
	] 
	table[x=iter, y=model, meta=portrait] {\eucoreDegPRHeatMap};
	
	\nextgroupplot[
		title=Chess,
		point meta min=0,
		point meta max=1,
		symbolic x coords={1,2,3,4,5,6,7,8,9,10},
		symbolic y coords={BTER,BUGGE,CL,CNRG,ER,GAE,GRNN,HRG,KRON,LAE,SBM},
		yticklabels={},
		xticklabel pos=top,
	]
	\addplot[
		matrix plot,
		draw=black, 
		fill opacity=1,
		mark=none,
		mesh/cols=10,
		point meta=explicit,
	] 
	table[x=iter, y=model, meta=portrait] {\chessDegPRHeatMap};
	
	\nextgroupplot[
		title=Cond-Mat,
		point meta min=0,
		point meta max=1,
		colorbar,
		colorbar style={width=5pt, xshift=0.15em, yshift=-1.5em, scale=0.8, transform shape},
		symbolic x coords={1,2,3,4,5,6,7,8,9,10},
		symbolic y coords={BTER,BUGGE,CL,CNRG,ER,GAE,GRNN,HRG,KRON,LAE,SBM},
		yticklabels={},
		xticklabel pos=top,
	]
	\addplot[
		matrix plot,
		draw=black, 
		fill opacity=1,
		mark=none,
		mesh/cols=10,
		point meta=explicit,
	] 
	table[x=iter, y=model, meta=portrait] {\condmatDegPRHeatMap};
	
	
	\nextgroupplot[
		ylabel={\textbf{$\lambda$-distance}},
		point meta min=-3,
		point meta max=1.7, 
		symbolic x coords={1,2,3,4,5,6,7,8,9,10},
		symbolic y coords={BTER,BUGGE,CL,CNRG,ER,GAE,GRNN,HRG,KRON,LAE,SBM},
	]
	\addplot[
		matrix plot,
		point meta min=-3,
		point meta max=1.7, 
		draw=black, 
		fill opacity=1,
		mark=none,
		mesh/cols=10,
		point meta=explicit,
	] 
	table[x=iter, y=model, meta expr=log10(\thisrow{lambda})] {\cliqueringDegPRHeatMap};	
	
	\nextgroupplot[
		point meta min=-3,
	   	point meta max=1.7, 
		symbolic x coords={1,2,3,4,5,6,7,8,9,10},
		symbolic y coords={BTER,BUGGE,CL,CNRG,ER,GAE,GRNN,HRG,KRON,LAE,SBM},
		yticklabels={},
	]
	\addplot[
		matrix plot,
		point meta min=-3,
		point meta max=1.7, 
		draw=black, 
		fill opacity=1,
		mark=none,
		mesh/cols=10,
		point meta=explicit,
	] 
	table[x=iter, y=model, meta expr=log10(\thisrow{lambda})] {\flightsDegPRHeatMap};	
	
	
	\nextgroupplot[
	   	point meta min=-3,
	   	point meta max=1.7, 
		symbolic x coords={1,2,3,4,5,6,7,8,9,10},
		symbolic y coords={BTER,BUGGE,CL,CNRG,ER,GAE,GRNN,HRG,KRON,LAE,SBM},
		yticklabels={},
	]
	\addplot[
		matrix plot,
		draw=black, 
		fill opacity=1,
		mark=none,
		mesh/cols=10,
		point meta=explicit,
	] 
	table[x=iter, y=model, meta expr=log10(\thisrow{lambda})] {\eucoreDegPRHeatMap};
	
	\nextgroupplot[
	   	point meta min=-3,
	   	point meta max=1.7, 
		symbolic x coords={1,2,3,4,5,6,7,8,9,10},
		symbolic y coords={BTER,BUGGE,CL,CNRG,ER,GAE,GRNN,HRG,KRON,LAE,SBM},
		yticklabels={},
	]
	\addplot[
		matrix plot,
		draw=black, 
		fill opacity=1,
		mark=none,
		mesh/cols=10,
		point meta=explicit,
	] 
	table[x=iter, y=model, meta expr=log10(\thisrow{lambda})] {\chessDegPRHeatMap};
	
	\nextgroupplot[
	    colorbar,
	    colorbar style={
	    	width=5pt, xshift=0.15em, yshift=-1.5em, scale=0.8, transform shape,
	    	ytick={-3,-1,1}, yticklabels={$10^{-3}$,$10^{-1}$,$10^1$},
	    },
	    point meta min=-3,
 		point meta max=1.7,
		symbolic x coords={1,2,3,4,5,6,7,8,9,10},
		symbolic y coords={BTER,BUGGE,CL,CNRG,ER,GAE,GRNN,HRG,KRON,LAE,SBM},
		yticklabels={},
	]
	\addplot[
		matrix plot,
		draw=black, 
		fill opacity=1,
		mark=none,
		mesh/cols=10,
		point meta=explicit,
	] 
	table[x=iter, y=model, meta expr=log10(\thisrow{lambda})] {\condmatDegPRHeatMap};

	\end{groupplot}
\end{tikzpicture}
    \caption{Portrait divergence (top) and $\lambda$-distance (bottom) values over 10 iterations on five real-world datasets averaged over 50 trials. Model failures are marked in gray. Confidence intervals are not plotted, but do not exceed $\pm$0.021 in the worst case. SBM and CNRG are the standout performers.}
    \label{fig:portrait}
\end{figure*}

\subsection{Edge Ratio and PageRank}
In this section, in the spirit of the previous Kronecker example, we will examine how the various models affect a graph's edge count and PageRank vector as the Infinity Mirror iterates.

One of the most obvious ways a model might distort a graph is by changing its total edge count.
Since some models do not make guarantees regarding a generated graph's edge count with respect to the input, measuring the raw edge count provides immediate insight into the change a graph might have undergone at the expense of some comprehensiveness.
With this in mind, we measure the change in edge count by tracking the number of edges in a generated graph in the Infinity Mirror sequence as a proportion of the number of edges in the original input graph.
Precisely, we define the \emph{edge ratio} by $\sfrac{|E(G_i)|}{|E(G_0)|}$ for a graph $G_i$ at iteration $i$ in the sequence compared to the initial $G_0$.

The first heatmap in Fig.~\ref{fig:degree} shows how the various models' edge ratios change across iterations on the five different datasets.
Shades more saturated in color (appearing darker) indicate a divergence from 1, indicating no loss, with shades closer to dark red indicating an increase and shades closer to dark blue representing a decrease.
The lighter shades closer to white indicate an edge ratio close to $1$, which would correspond to very little if any change in the edge count.
The grey rows indicate data missing either due to models failing to fit/generate on the input dataset.

As we can see, most models behave quite well as measured by the edge ratio: ER, BTER, SBM, CL, and CNRG all have an edge ratio close to $1$, which is expected from the way these models are defined.
Since ER is a random graph model whose connection parameter is learned from the input, the expected edge count for any graph generated by that model should be the same as the input's edge count.
Similar reasoning extends to BTER and SBM; since their probabilities are learned from the input, the number of expected edges in any generated graph should be similar to the original.
Further, since CL attempts to replicate the input graph's degree distribution, any generated graph should have nearly the same number of edges as the original.

The prominent deviants in Fig.~\ref{fig:degree} are BUGGE, KRON, and the graph neural network models.
On three of the five datasets, BUGGE generates graphs markedly sparser than the original.
This is likely a consequence of the nature of the Kemp-Tenenbaum grammars that BUGGE learns not being well-suited to capturing global information regarding the raw edge count of the graph.
The grammars learned by BUGGE are also unable to capture the fine structure of the already-sparse Clique-Ring dataset, leading to a slight increase in the edge count.
As expected from the analysis in Fig~\ref{fig:kronexample}, KRON exhibits an exponential increase in the number of edges across all datasets as the model is successively refitted on its outputs. This is shown clearly by the linear color gradient in Fig~\ref{fig:degree}, since the colors for this particular heatmap are scaled logarithmically.

The two graph autoencoders -- GAE and LAE -- remarkably increase the edge counts of their generated graphs from as early as the first iteration of the Infinity Mirror test.
We can see across all five datasets that GAE and LAE compare only to the most mature iterations of KRON.
This immediate and sustained increase in density is unique to the two autoencoders and foreshadows their characteristic behavior on the other graph similarity metrics.
Finally, we can observe a relatively modest increase in edge count by GRNN on Flights, the smallest of the real-world datasets, contrasted by a drastic decrease in edge count after the first iteration on Clique-Ring.
Since GRNN seems to behave comparably well on Eucore, it may be too early for any conclusory remarks regarding GRNN.

The second heatmap in Fig.~\ref{fig:degree} displays the Jensen-Shannon divergence of the models' PageRank distributions between $G_i$ and $G_0$. The PageRank algorithm assigns a weight to each node PR$(v)$, signifying a random walker's probability over the graph's edges landing on that node. A node's PageRank score is highly correlated to its degree but contains additional information about the graph's topology. However, since PageRank scores may not necessarily be integers, the JS divergence cannot be computed directly.
In our analysis we create $100$ equal-length bins from $0$ to $\underset{v \in \{V_0\cup V_i\}}{\max}\textrm{PR}(v)$ and compute the JS divergence on these binned PageRank distributions.

It is important to note three trends from this second heatmap that will remain largely true throughout the rest of the analyses:
\begin{itemize}
\item KRON gradually degrades as iterations increase,
\item
    GAE and LAE immediately diverge across datasets, and
\item CNRG performs abnormally well on the Clique-Ring dataset due to the Vertex Replacement Grammar's ability to perfectly replicate the dataset's intricate clustering.
\end{itemize}
Additionally, we can see that GRNN performs about as poorly as the random ER model on the two real-world datasets for which data is available.
Finally, none of the models performs particularly well on Clique-Ring with the aforementioned exception of CNRG.


\subsection{Portrait Divergence and $\lambda$-Distance}
Edge ratio and PageRank both describe a graph from an important but limited perspective. 
Clearly, two graphs with the same number of edges may have drastically different topologies.
Similarly, graphs with similar PageRank distributions do not necessarily share similar network topology~\citep{prvzulj2007biological}.
A more thorough examination that compares the topology and internal substructures between the source graph and each new iteration is needed to describe each model in more detail. 
Among the many options~\citep{tantardini2019comparing}, we focus on Portrait divergence~\citep{bagrow2019information} and $\lambda$-distance~\citep{wilson2008study}.

The Portrait divergence is based on a \textit{portrait} of a graph~\citep{bagrow2008portraits}, based on the distribution of pairwise shortest path lengths. It is represented by matrix $B = [b_{\ell, k}]$, the number of nodes which are exactly $\ell$ hops away from $k$ other nodes, where $\ell \in [0, d]$, where $d$ is the graph diameter, and $k \in [0, |V|)$. Because shortest path lengths are correlated with edge density, degree distribution, and other centrality measures, the network portrait is a comprehensive (and often visually compelling) summary of a graph's topology. Furthermore, the graph portrait provides a probability $P(k,\ell)$ of randomly choosing exactly $k$ nodes at a distance $\ell$. It, therefore, provides a probability distribution over all nodes in a graph. The Portrait divergence is therefore defined as the JS divergence of portrait distributions from two graphs~\citep{bagrow2019information}.

\begin{figure}
    \centering
    \scalebox{0.68}{
\setlength{\tabcolsep}{0.44em}
\centering
\begin{tabular}{@{}l r rrrrrrrrr@{}}
\toprule
&
\multicolumn{1}{c}{
\begin{tikzpicture}[scale=0.5, transform shape]
\node [inode, outer sep=1pt] (v1) at (-4.25,4) {};
\node [inode, outer sep=1pt] (v2) at (-3.5,4) {};
\node [hidden] at (-4.25,3.25) {};
\node [hidden] at (-3.5,3.25) {};

\draw [edge] (v1) edge (v2);

\end{tikzpicture}
} &
\multicolumn{1}{c}{
\begin{tikzpicture}[scale=0.5, transform shape]
\node [inode, outer sep=1pt] (v1) at (-4.25,4) {};
\node [inode, outer sep=1pt] (v2) at (-3.5,4) {};
\node [inode, outer sep=1pt] (v3) at (-4.25,3.25) {};
\node [hidden] (v4) at (-3.5,3.25) {};

\draw [edge] (v1) edge (v2);
\draw [edge] (v1) edge (v3);
\end{tikzpicture} 
} &
\multicolumn{1}{c}{

\begin{tikzpicture}[scale=0.5, transform shape]
\node [inode, outer sep=1pt] (v1) at (-4.25,4) {};
\node [inode, outer sep=1pt] (v2) at (-3.5,4) {};
\node [inode, outer sep=1pt] (v3) at (-4.25,3.25) {};
\node [hidden] (v4) at (-3.5,3.25) {};

\draw [edge] (v1) edge (v2);
\draw [edge] (v1) edge (v3);
\draw [edge] (v2) edge (v3);
\end{tikzpicture} 
} &
\multicolumn{1}{c}{
\begin{tikzpicture}[scale=0.5, transform shape]
\node [inode, outer sep=1pt] (v1) at (-4.25,4) {};
\node [inode, outer sep=1pt] (v2) at (-3.5,4) {};
\node [inode, outer sep=1pt] (v3) at (-4.25,3.25) {};
\node [inode, outer sep=1pt] (v4) at (-3.5,3.25) {};

\draw [edge] (v1) edge (v2);
\draw [edge] (v2) edge (v4);
\draw [edge] (v1) edge (v3);
\end{tikzpicture} 
} &
\multicolumn{1}{c}{
\begin{tikzpicture}[scale=0.5, transform shape]
\node [inode, outer sep=1pt] (v1) at (-4.25,4) {};
\node [inode, outer sep=1pt] (v2) at (-3.5,4) {};
\node [inode, outer sep=1pt] (v3) at (-4.25,3.25) {};
\node [inode, outer sep=1pt] (v4) at (-3.5,3.25) {};

\draw [edge] (v1) edge (v2);
\draw [edge] (v1) edge (v3);
\draw [edge] (v1) edge (v4);
\end{tikzpicture} 
} &
\multicolumn{1}{c}{
\begin{tikzpicture}[scale=0.5, transform shape]
\node [inode, outer sep=1pt] (v1) at (-4.25,4) {};
\node [inode, outer sep=1pt] (v2) at (-3.5,4) {};
\node [inode, outer sep=1pt] (v3) at (-4.25,3.25) {};
\node [inode, outer sep=1pt] (v4) at (-3.5,3.25) {};

\draw [edge] (v1) edge (v2);
\draw [edge] (v2) edge (v4);
\draw [edge] (v1) edge (v3);
\draw [edge] (v3) edge (v4);
\end{tikzpicture}
} &
\multicolumn{1}{c}{
\begin{tikzpicture}[scale=0.5, transform shape]
\node [inode, outer sep=1pt] (v1) at (-4.25,4) {};
\node [inode, outer sep=1pt] (v2) at (-3.5,4) {};
\node [inode, outer sep=1pt] (v3) at (-4.25,3.25) {};
\node [inode, outer sep=1pt] (v4) at (-3.5,3.25) {};

\draw [edge] (v1) edge (v2);
\draw [edge] (v2) edge (v4);
\draw [edge] (v1) edge (v4);
\draw [edge] (v3) edge (v1);
\end{tikzpicture} 
} &
\multicolumn{1}{c}{
\begin{tikzpicture}[scale=0.5, transform shape]
\node [inode, outer sep=1pt] (v1) at (-4.25,4) {};
\node [inode, outer sep=1pt] (v2) at (-3.5,4) {};
\node [inode, outer sep=1pt] (v3) at (-4.25,3.25) {};
\node [inode, outer sep=1pt] (v4) at (-3.5,3.25) {};

\draw [edge] (v1) edge (v2);
\draw [edge] (v2) edge (v4);
\draw [edge] (v1) edge (v4);
\draw [edge] (v3) edge (v1);
\draw [edge] (v3) edge (v4);
\end{tikzpicture} 
}&
\multicolumn{1}{c}{
\begin{tikzpicture}[scale=0.5, transform shape]
\node [inode, outer sep=1pt] (v1) at (-4.25,4) {};
\node [inode, outer sep=1pt] (v2) at (-3.5,4) {};
\node [inode, outer sep=1pt] (v3) at (-4.25,3.25) {};
\node [inode, outer sep=1pt] (v4) at (-3.5,3.25) {};

\draw [edge] (v1) edge (v2);
\draw [edge] (v2) edge (v4);
\draw [edge] (v1) edge (v4);
\draw [edge] (v3) edge (v1);
\draw [edge] (v3) edge (v4);
\draw [edge] (v2) edge (v3);
\end{tikzpicture} 
} \\ 

& \multicolumn{1}{c}{\scriptsize{\textsf{edge}}} & \multicolumn{1}{c}{\scriptsize{$2$-\textsf{star}}} & \multicolumn{1}{c}{\scriptsize{\textsf{triangle}}} & \multicolumn{1}{c}{\scriptsize{$4$-\textsf{path}}} & \multicolumn{1}{c}{\scriptsize{$3$-\textsf{star}}} & \multicolumn{1}{c}{\scriptsize{$4$-\textsf{cycle}}} &  \multicolumn{1}{c}{\scriptsize{$4$-\textsf{tail-tri}}} & \multicolumn{1}{c}{\scriptsize{$4$-\textsf{chord-cyc}}}  & \multicolumn{1}{c}{\scriptsize{$4$-\textsf{clique}}} \\
\midrule
$\textsf{PCA}_x$ &  \greenpos{0.034} & \redneg{$-$0.017} &   \greenpos{0.031} & \redneg{-0.603} & \redneg{-0.394} &   \greenpos{0.009} &         \greenpos{0.256} &        \greenpos{0.182} &    \greenpos{0.617} \\
$\textsf{PCA}_y$ & \redneg{-0.223} & \redneg{-0.230} & \redneg{-0.123} & \redneg{-0.481} &  \greenpos{0.606} &  \redneg{-0.018} &        \redneg{-0.481} &       \redneg{-0.147} &    \greenpos{0.173} \\

\bottomrule
\end{tabular}
}
    \vspace*{.2cm}
    \input{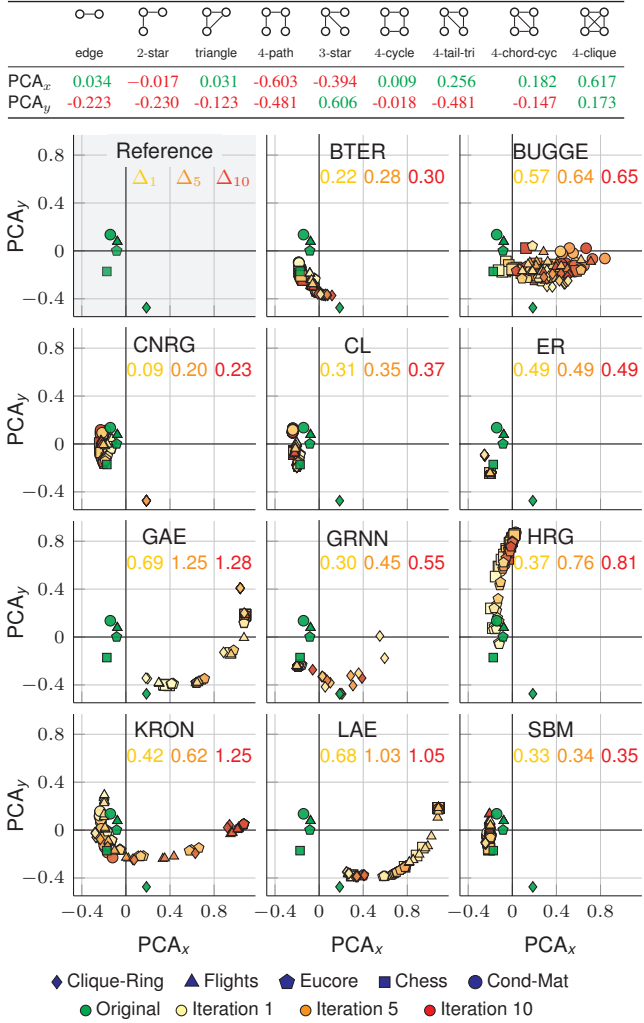}
    \vspace*{.2cm}
    \begin{tikzpicture}
    \begin{customlegend}[ 
    legend columns=5,
    legend style={
    draw=none,
    font=\scriptsize,
  },
  legend cell align={left},
  legend entries={\textsf{Clique-Ring}~~~~,  \textsf{Flights}~~~~, \textsf{Eucore}~~~~, \textsf{Chess}~~~~, \textsf{Cond-Mat}~~~~}
  ]
    \addlegendimage{tree,  mark options={solid, scale=1.25, fill=blue}, draw=black,  draw opacity=1, fill opacity=1}
    \addlegendimage{flights, mark options={solid, scale=1.5, fill=blue}, draw=black, draw opacity=1, fill opacity=1}
    \addlegendimage{eucore, mark options={solid, scale=1.5, fill=blue}, draw=black,  draw opacity=1, fill opacity=1}
    \addlegendimage{chess, mark options={solid, scale=1, fill=blue}, draw=black,  draw opacity=1, fill opacity=1}
    \addlegendimage{cliquering, mark options={solid, scale=1.25, fill=blue}, draw=black,  draw opacity=1, fill opacity=1}
    \end{customlegend}
\end{tikzpicture}

\vspace*{-1em}
\begin{tikzpicture}
    \begin{customlegend}[ 
    legend columns=4,
    legend style={
    draw=none,
    font=\scriptsize,
  },
  legend cell align={left},
  legend entries={\textsf{Original}~~~, \textsf{Iteration 1}~~~~~, \textsf{Iteration 5}~~~~~, \textsf{Iteration 10}~~~~} 
  ]
    \addlegendimage{only marks, mark=*, mark options={solid, fill=green}}
    
    \addlegendimage{only marks, mark=*, mark options={solid, fill=yellow!50}}
    \addlegendimage{only marks, mark=*, mark options={solid, fill=orange}}
    \addlegendimage{only marks, mark=*, mark options={solid, fill=red}}
    \end{customlegend}
\end{tikzpicture}
    \caption{PCA weights for the graphlet vectors (top). 2-D PCA of graphlet vectors (bottom) on all five datasets (represented by shape) and all 11 graph models (represented by separate plots), showing the first 10 iterations (represented by color). The original graphs are plotted in green. Each mark represents the coordinate of a generated graph averaged over all trials. These illustrations show how graphs degenerate comparatively over multiple iterations. The numbers $\Delta_1$, $\Delta_5$, and $\Delta_{10}$ in each plot represent the average Euclidean distance between the embeddings of the original and generated graphs for iterations 1, 5, and 10 respectively.}
    \label{fig:pca}
\end{figure}

The $\lambda$-distance is similar to Portrait divergence. It is defined as the Euclidean distance between the eigenvalues of two graphs. In the present work, we use the graph Laplacian $\mathcal{L} = D - A$ (\ie, the difference between the degree and adjacency matrices) instead of the adjacency matrix due to its desirable properties (\eg, lower co-spectrality between non-isomorphic graphs)~\citep{wilson2008study}. 

Results of Portrait divergence and $\lambda$-distance are plotted as heatmaps in Fig.~\ref{fig:portrait}. 
Like in Fig.~\ref{fig:degree}, each block represents the mean over 50 trials on each source graph and model over the first 10 iterations. 
Again, confidence intervals are not plotted for clarity but do not exceed $\pm$0.021 in the worst case. 
By design, the Infinity Mirror test looks to accumulate and compound the \textit{errors} and the biases in the generated graphs. This is confirmed in both Figs.~\ref{fig:degree} and \ref{fig:portrait} where the colors of the cells change from yellow to red. 

We want to draw the reader's attention to the performance of BTER, BUGGE, and CL models. While they are all able to capture and mimic the first-order distributions like PageRank, evidenced by the blue cells in the heatmaps in Fig.~\ref{fig:degree}, they struggle with higher-order distances like Portrait divergence.
Understanding the reasons why that is the case is important. We hypothesize that this is because these models are not sensitive enough to detect higher-order edge mixing patterns. CL, for example, only tracks the degree distribution of its input. BTER goes a step further and augments that with average clustering information. BUGGE's grammar rules are designed for directed graphs and are optimized to be small in size, as a result of which it sometimes fails to capture mesoscale properties. These observations are not meant to criticize these models but to understand their behaviors better.
In comparison, CNRG and SBM demonstrate their prowess by achieving low scores across both first-order and higher-order measures. Their performances warrant praise, but we continue our investigation using more sophisticated tools. We start by moving our focus from edges and paths to local structures and tracking triadic closures.

\subsection{Tracking Degeneracy: Graphlets}
Graph modelers have also found that the number and types of various graphlets significantly contribute to the graph's overall topology. 
For every graph, we use the Parallel Parameterized Graphlet Decomposition (PGD) algorithm~\citep{ahmed2015efficient} to count occurrences of 2, 3, and 4-node connected subgraphs and store the result in a 9-dimensional embedding vector. These vectors are $L_2$ normalized and then reduced with principal component analysis (PCA) into 2-dimensions. 
Fig.~\ref{fig:pca} show the results of PCA on the graphlet vectors with each model separated. Therefore, each mark in the plot for a specific model represents the reduced vector, averaged over 50 trials for each iteration. Different datasets are represented with different shapes, while different iterations follow a gradient of yellow to red from iteration 1 to 10. The coordinates of the original datasets $G_0$ are marked in green to serve as a reference. 
To quantify the dispersion from the original graph, we compute the Euclidean distance between the embeddings of the original graphs and the generated graphs. We aggregate these distances for each dataset model pair and report their means for iterations 1, 5, and 10 in each of the plots in Fig.~\ref{fig:pca}. 

PCA was used instead of t-SNE or other dimension reduction techniques because its reduced vectors are a simple linear combination of weights on the original vector. This provides a somewhat understandable representation for each reduced vector if the original vector also carries semantic meaning. The top part of Fig.~\ref{fig:pca} shows the weights for each element that, when combined, map each mark to its 2-dimensional point. Interpreting PCA weights is fraught with difficulty, but in this case, the findings are pretty straightforward: the $x$-axis is positively correlated with 4-clique and 4-tailed-triangle and negatively correlated with 4-path and 3-star, while the $y$-axis is highly correlated with 3-star, negatively correlated with 4-tailed-triangles and 4-paths. 

Representing axes in this way allows the reader to begin to conclude the plot in Fig.~\ref{fig:pca}.
We call-out two interesting findings as examples. First, from the HRG plot,  we observe that the markers shift upwards in the $y$-axis with increasing iterations, as evidenced by the change in marker colors from yellow to red. This could be due to an increase in the number of $4$-cliques and $3$-stars.  Simply put, the HRG model appears to be biased towards generating more cliques. The observation can validate that the \textit{hyperedges} in the HRG model are formed directly from a clique-tree. 

Second, the graph autoencoder models (GAE and LAE) and KRON tend to diverge quickly from the input graphs, away from all the models, to occupy a region with high values of $x$. High $x$ values could result from a large number of $4$-cliques and a lack of $4$-paths  and $3$-stars, indicating an increase in the density of edges in the generated graphs. This transition is gradual, as seen by both the trail of markers changing colors from yellow to red and the increasing distances. 

Third, CNRG, SBM, BTER, and CL perform similarly well. The markers overlap closely with the originals', evidenced by the low distance values. Indeed CL's performance is surprising, particularly for the four real-world graphs. 
These prompted us to examine the topological degeneracy of the graphs from yet another perspective in Sec.~\ref{sec:aplcc}, looking at how average clustering and average shortest path lengths evolve across iterations for different models. 
There are certainly additional conclusions that may be drawn from the PCA plots, but we leave these for the reader. 

\begin{figure}[tb]
    \centering
    \input{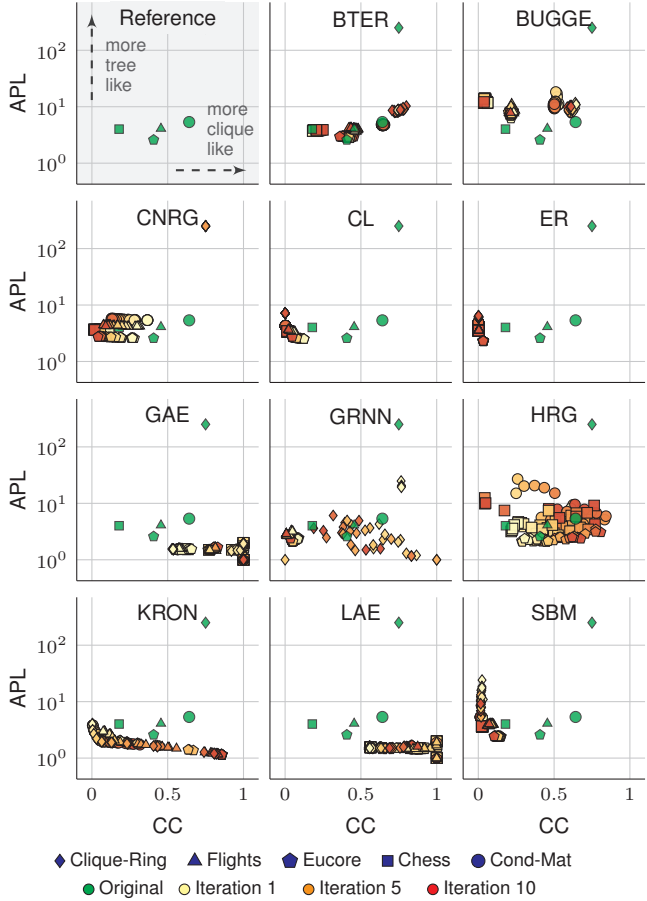}
    \vspace*{.2cm}
    \begin{tikzpicture}
    \begin{customlegend}[ 
    legend columns=5,
    legend style={
    draw=none,
    font=\scriptsize,
  },
  legend cell align={left},
  legend entries={\textsf{Clique-Ring}~~~~,  \textsf{Flights}~~~~, \textsf{Eucore}~~~~, \textsf{Chess}~~~~, \textsf{Cond-Mat}~~~~}
  ]
    \addlegendimage{tree,  mark options={solid, scale=1.25, fill=blue}, draw=black,  draw opacity=1, fill opacity=1}
    \addlegendimage{flights, mark options={solid, scale=1.5, fill=blue}, draw=black, draw opacity=1, fill opacity=1}
    \addlegendimage{eucore, mark options={solid, scale=1.5, fill=blue}, draw=black,  draw opacity=1, fill opacity=1}
    \addlegendimage{chess, mark options={solid, scale=1, fill=blue}, draw=black,  draw opacity=1, fill opacity=1}
    \addlegendimage{cliquering, mark options={solid, scale=1.25, fill=blue}, draw=black,  draw opacity=1, fill opacity=1}
    \end{customlegend}
\end{tikzpicture}

\vspace*{-1em}
\begin{tikzpicture}
    \begin{customlegend}[ 
    legend columns=4,
    legend style={
    draw=none,
    font=\scriptsize,
  },
  legend cell align={left},
  legend entries={\textsf{Original}~~~, \textsf{Iteration 1}~~~~~, \textsf{Iteration 5}~~~~~, \textsf{Iteration 10}~~~~} 
  ]
    \addlegendimage{only marks, mark=*, mark options={solid, fill=green}}
    
    \addlegendimage{only marks, mark=*, mark options={solid, fill=yellow!50}}
    \addlegendimage{only marks, mark=*, mark options={solid, fill=orange}}
    \addlegendimage{only marks, mark=*, mark options={solid, fill=red}}
    \end{customlegend}
\end{tikzpicture}
    \caption{Variation of average clustering coefficient (CC) and average path length (APL) across all five datasets (represented by shape) and 11 graph models (represented as separate plots), showing all 10 iterations (represented by color). Each mark represents the coordinate of a generated graph averaged over all trials. The reference plot on the top left is included to help the reader interpret the plots.}
    \label{fig:aplcc}
\end{figure}

\subsection{Tracking Degeneracy: Clustering and Geodesics} \label{sec:aplcc}

Yet another way to track the topological changes in graphs throughout Infinity Mirror generations is tracking changes in clustering and shortest path lengths. The average path length (APL) and average clustering coefficient (CC) measures capture important topological information about a graph and provide a meaningful summarization of a graph's shape and structure. 

Average clustering measures the amount of triadic closure in a graph, roughly the fraction of closed triangles. Real-world social networks often contain a lot more triangles, and therefore have higher average clustering compared to ER graphs~\citep{klimek2013triadic}. 

Similarly, distances between nodes, as measured by the lengths of shortest paths, indicate how far apart or close together the nodes in a graph are, with meaningful implications in many applications.
Graphs mined from real-world data, such as social and biological networks, often exhibit a low average shortest path length, known as the \emph{small world} property~\citep{watts1998collective}. Synthetic graphs (\eg, trees, cycles), on the other hand, can have much longer geodesics.
If a graph model hopes to capture the global nature of interactions in a network, then the average path length is an important property to preserve. Tab.~\ref{tab:dataset} has the average clustering, average shortest path lengths used in the paper. 

In Fig.~\ref{fig:aplcc}, we have plotted the average clustering (CC) against the average path length (APL) of eleven generative models on five datasets as they experience 50 iterations of the Infinity Mirror test. After the 10th iteration, the results tend to stabilize (or completely degenerate), so we only plot the first 10 in most illustrated results.
Points plotted are aggregated means across the 50 independent chains of the Infinity Mirror, with the initial graph $G_0$ for each dataset colored green and subsequent iterations $G_1$ through $G_{10}$ colored in a gradient from yellow to red.
Different shaped markers distinguish datasets.

Similar to Fig.~\ref{fig:pca}, KRON, GAE, and LAE display a similarly gradual progression, whereby the APL values consistently and monotonically decrease as the clustering coefficients creep higher over the iterations.
This further validates our previous observation that these models create highly dense graphs as the model is iteratively fit.

The ER and CL graphs display a similarly sudden and consistent grouping of points towards the lower left of the graph. This corresponds with the common knowledge that these random graph models have low clustering and small inter-node distances.

CNRG and BTER perform markedly well on these metrics. We can see that the points for each dataset stay reasonably near their initial starting point for a few iterations before gradually tending to decrease clustering.
On all datasets, either CC or APL is very well preserved, while the other metric slowly drifts away from the initial values.

Interestingly, the BUGGE graphs can maintain the same average clustering while getting less tree-like, \ie, denser. This could be due to the introduction of edges across distant regions, which perform as random shortcuts, bringing down the APL without affecting the CC. 

Finally, GRNN displays peculiar behavior.
GRNN displays neither the tight, sudden grouping nor the gradual, consistent decay of some other models.
Instead, the points seem to generally congregate near the bottom of the plot, indicating a consistent decrease in APL but inconsistent behavior in terms of clustering.

\subsection{Investigating Learned Model Spaces}

\begin{figure*}
    \centering
    \input{figures/model-study.tex}
    \caption{Model parameters for iterations 1, 5, 10, 20, and 50 on the Ring of Cliques graph using KRON, SBM, BUGGE, and CNRG models. Kronecker initiator matrices tend towards denser graphs. The SBM model tends towards fewer, sparser communities. Higher probabilities are redder, and only the non-zero elements on the block matrices are drawn. We only draw the right-hand side of the two most frequent BUGGE production rules, which show that a $K_5$ is encoded (instead of a $K_4$ from the source graph). CNRG generates graphs that are isomorphic to the source graph where $\bar{C}_{500}$ is a 500 node cycle. Rule frequencies are denoted by $\times$.}
    \label{fig:models}
\end{figure*}

Graph metrics like edge ratios and PageRank distribution from Fig.~\ref{fig:degree}, Portrait and $\lambda$-distance from Fig.~\ref{fig:portrait}, generally show that models tend to degenerate as they are repeatedly asked to model their generated graph. The result from Fig.~\ref{fig:pca} offers clues to how they degenerate, but these plots investigate the output of the models, not the model itself. 

Next, we investigate how the models change to gain further insights into the biases encoded into each model. Unfortunately, only a handful of the graph models studied in the present work contain interpretable parameters. For example, the model parameters of neural networks (\ie, GRNN, LAE, GAE, and NGAN) are well known to be challenging to interpret. The BTER, HRG, and CL parameters carry semantic meaning but are still difficult for humans to interpret. Likewise, the ER model is just the number of nodes and edges, which do not degenerate across iterations. What remains are the KRON, SBM, BUGGE, and CNRG models. 

We display the interpretable graph model parameters $\Theta$ for iterations 1, 5, 10, 20, and 50 of the Clique-Ring source graph in Fig.~\ref{fig:models}. 
The KRON model shows the initiator matrix, which generates a graph by repeatedly computing the Kronecker product and stochastically drawing an edge in the resulting matrix according to the value in each cell. We find that all four entries in the initiator matrix tend towards 0.999. This leads to the generation of increasingly dense graphs, similar to what we observed in Fig.~\ref{fig:kronexample} and Fig.~\ref{fig:aplcc}. 

The SBM detects and models graphs using a reduced block matrix. New graphs are generated by drawing tight-knit groups according to the probabilities in each block and connecting disparate communities according to their off-diagonal probabilities. In $\Theta_1$, SBM discovers the ring structure, evidenced by most communities being only connected to two neighboring communities. 
We see that, as the iterations increase, the number of detected communities decreases. The matrices also start resembling a core-periphery structure, the cores consisting of many nodes, with the rest appearing in sparse communities surrounding the core.

Node replacement graph grammar models of the Clique-Ring produced by BUGGE and CNRG are easily interpretable. These models encode context-free grammars of graph substructures where a node represented on the left-hand side of a production rule is replaced with the graph on the rule's right-hand side. CNRG first encodes the individual 4-cliques into separate nonterminal nodes (bottom rule) and then encodes the 500-node ring into a single starting rule. In this way, CNRG can capture the \textit{exact} topology of the graph and is, therefore, able to generate an isomorphic copy over multiple iterations.

BUGGE treats all edges as bi-directed, including in the production rules, which show an interesting bias. In the first iteration $\Theta_1$, the top rule encodes cliques of arbitrary size, and the bottom rule encodes a chain of cliques. These rules still preserve the clique-ring structure, but the size of the individual cliques starts to vary. Because the top rule in the first iteration does not limit either the number or the size of individual cliques, we observe that 5-cliques start frequently appearing in later iterations.

\subsection{Running Time Analysis}
These results show that the Infinity Mirror test can expose interesting degeneration processes of graph models. While this is a useful outcome, the test's efficacy must also consider the cost of performing such a test. This section presents and analyzes timings associated with the included models and the Infinity Mirror test.
    
\begin{figure}
    \centering
    \input{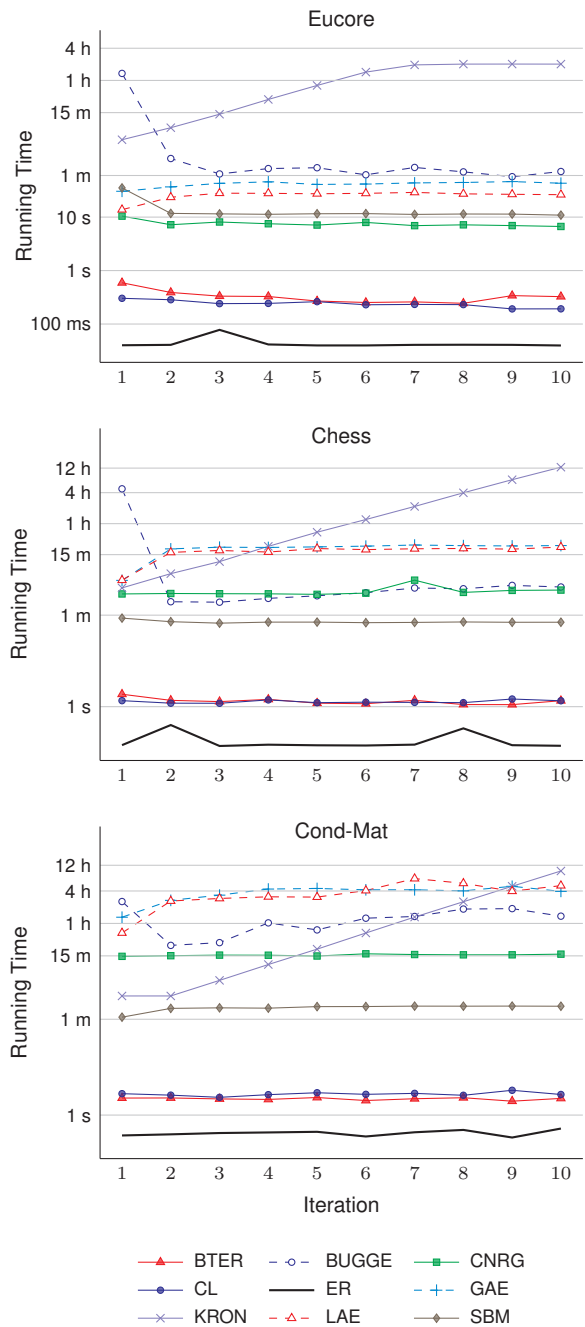}

\begin{tikzpicture}[font=\sffamily]
\begin{groupplot}[
        group style={
            group size=1 by 3, 
            group name=myplot,
            horizontal sep=0pt,
            vertical sep=25pt,
            xlabels at=edge bottom,
            ylabels at=edge left,
        },
    ylabel style={align=center, yshift=0.05em},
    ymajorgrids,
    height=6cm,
    width=8cm,
    axis lines*=left,
    x tick style=transparent,
    y tick style=transparent,
    axis on top,
    enlarge x limits=0.05,
    enlarge y limits=0.15,
    legend columns=3,
    legend cell align=left,
    legend style={font=\scriptsize,
    draw=none, at={(0.5,-2.65)},anchor=north},
    xmin=1,
    xmax=10,
    ymode=log,
    tick label style={font=\scriptsize}, 
    ylabel={\footnotesize{\textsf{Running Time}}},
    xlabel={\footnotesize{\textsf{Iteration}}},
    xtick={1,2,...,10},
    ytick={0.1,1,10,60,900,3600,14400},
    yticklabels={100 ms,1 s,10 s,1 m,15 m,1 h,4 h},
    minor tick num=1,
    xtick pos=bottom, 
    ytick pos=left,
]


\nextgroupplot[
    title={\footnotesize{Eucore}}, 
    ylabel style={align=center, yshift=-0.98em},
    title style={yshift=-8pt,}, 
    ymin=0.1,
    ymax=6000
]

    \addplot [bter] table [
        x=iter, y=total_time
        ] 
    {\eucoreBTER}; 
    \addlegendentry{~BTER~~~~};
     
    \addplot [bugge] table [
        x=iter, y=total_time
        ] 
    {\eucoreBUGGE};  
    \addlegendentry{~BUGGE~~~~};
    
    \addplot [cnrg] table [
        x=iter, y=total_time
        ] 
    {\eucoreCNRG};  
    \addlegendentry{~CNRG~~~~};

    \addplot [cl] table [
        x=iter, y=total_time
        ] 
    {\eucoreCL};  
    \addlegendentry{~CL~~~~};

    \addplot [er] table [
        x=iter, y=total_time
        ] 
    {\eucoreER};  
    \addlegendentry{~ER~~~~};

    \addplot [gcnae] table [
        x=iter, y=total_time
        ] 
    {\eucoreGAE}; 
    \addlegendentry{~GAE~~~~};

    \addplot [kron] table [
        x=iter, y=total_time
        ] 
    {\eucoreKRON};  
    \addlegendentry{~KRON~~~~};

    \addplot [linae] table [
        x=iter, y=total_time
        ] 
    {\eucoreLAE};  
    \addlegendentry{~LAE~~~~};

    \addplot [sbm] table [
        x=iter, y=total_time
        ] 
    {\eucoreSBM};  
    \addlegendentry{~SBM~~~~};

\nextgroupplot[
    title={\footnotesize{Chess}}, 
    title style={yshift=-15pt,}, 
     ytick={1,60,900,3600,14400,43200},
    yticklabels={1 s,1 m,15 m,1 h,4 h,12 h},
    ymin=0.5,
    ymax=46000
]

    \addplot [bter] table [
        x=iter, y=total_time
        ] 
    {\chessBTER}; 
     
    \addplot [bugge] table [
        x=iter, y=total_time
        ] 
    {\chessBUGGE};  
    
    \addplot [cnrg] table [
        x=iter, y=total_time
        ] 
    {\chessCNRG};  
    
    \addplot [cl] table [
        x=iter, y=total_time
        ] 
    {\chessCL};  
    
    \addplot [er] table [
        x=iter, y=total_time
        ] 
    {\chessER};  
    
    \addplot [gcnae] table [
        x=iter, y=total_time
        ] 
    {\chessGAE};  
    
    \addplot [kron] table [
        x=iter, y=total_time
        ] 
    {\chessKRON};  
    
    \addplot [linae] table [
        x=iter, y=total_time
        ] 
    {\chessLAE};  
    
    \addplot [sbm] table [
        x=iter, y=total_time
        ] 
    {\chessSBM};

\nextgroupplot[
    title={\footnotesize{Cond-Mat}}, 
    title style={yshift=-14pt,}, 
    ytick={1,60,900,3600,14400,43200},
    yticklabels={1 s,1 m,15 m,1 h,4 h,12 h},
    ymin=0.8,
    ymax=44000
]

    \addplot [bter] table [
        x=iter, y=total_time
        ] 
    {\CondMatBTER}; 
     
    \addplot [bugge] table [
        x=iter, y=total_time
        ] 
    {\CondMatBUGGE};  
    
    \addplot [cnrg] table [
        x=iter, y=total_time
        ] 
    {\CondMatCNRG};  
    
    \addplot [cl] table [
        x=iter, y=total_time
        ] 
    {\CondMatCL};  
    
    \addplot [er] table [
        x=iter, y=total_time
        ] 
    {\CondMatER};  
    
    \addplot [gcnae] table [
        x=iter, y=total_time
        ] 
    {\CondMatGAE};  
    
    \addplot [kron] table [
        x=iter, y=total_time
        ] 
    {\CondMatKRON};  
    
    \addplot [linae] table [
        x=iter, y=total_time
        ] 
    {\CondMatLAE};  
    
    \addplot [sbm] table [
        x=iter, y=total_time
        ] 
    {\CondMatSBM};

\end{groupplot}
\end{tikzpicture}
    \caption{Running time for 10 iterations of models for the three largest datasets averaged over 50 trials. Most models take roughly the same amount of time across iterations with the exception of KRON and BUGGE.}
    \label{fig:runtime}
\end{figure}

Ultimately, the cost of the Infinity Mirror test is a summation of the cost of running multiple generate-and-fit iterations. 
We report running times over 10 iterations of the test for all models on the three largest graphs in Fig.~\ref{fig:runtime} averaged over 50 independent trials. 
As expected, running time is proportional to the size of the current input graph. Most models, except KRON and BUGGE, as observed in Fig.~\ref{fig:degree}(A), maintain a constant edge ratio, therefore, have roughly the same running time across iterations.
KRON's tendency to increasingly densify the graph causes the runtime to scale exponentially.

BUGGE's timing curves reveal an interesting pattern: the first iteration takes significantly longer than the rest.
In addition to the dimensions of the graph, BUGGE's running time also depends on the degree of structure present in the graph.
Real-world graphs have complicated, non-uniform local structures that significantly increase the search space for possible grammar rules. At the end of the first iteration, BUGGE finds a simple, interpretable set of grammar rules to describe the original graph, then used to generate future graphs.
However, these new graphs are highly structured and lack the local structural nuances of the original input graph. So, successive runs of BUGGE become considerably faster. 

Finally, we acknowledge that our analysis does not include the cost of computing the various distributions. 
First-order metrics like degree, PageRank, and average clustering can be computed efficiently and add very little overhead. 
Graphlet counting, $\lambda$-distance, and Portrait divergence computations, on the other hand, are more expensive and do not scale as well, sometimes taking several hours to finish for dense graphs.

\section{Conclusions} \label{sec:findings} 

To conclude, we combine the observations from the experiments and discuss our findings in a broader context.
We find that the graph neural network-based models failed to perform well in our test by not capturing the basic graph properties. The autoencoders LAE and GAE perform admirably in link prediction tasks. However, we demonstrated that graphs generated from their embeddings share surprisingly little topological similarity with the input. We also unveiled scalability and consistency issues with GRNN, which was unexpected since it is designed to learn structural patterns.

For some models, the biases are somewhat obfuscated. However, the test comes through and informs us that CL, HRG, KRON, and BTER are good at capturing specific properties but not so good at others. 
This is particularly true for BTER and CL, which target specific distributions from the input graph during the fitting process.
Similarly, the test highlights scalability and stability issues with HRGs, which we suspect results from its dependence on tree decompositions.

SBM and CNRG perform admirably on the Infinity Mirror test. For SBM, we suspect this stems from its ability to model and capture node interaction patterns on a mesoscopic level. CNRG, on the other hand, goes a step further by leveraging and modeling the natural hierarchy of nodes in the graphs.

In summary, this work presents a new methodology for analyzing graph models by repeatedly fitting and generating.
These iterative fit-and-generate processes can reveal hidden model biases, facilitating their critical examination.

Finally, we hope that these findings will be used to more deeply understand the statistical biases encoded into graph models and aid in developing more robust graph models.

\section*{Acknowledgements} 
This work is supported by grants from the US National Science Foundation (\#1652492), DARPA (FA8750-20-2-1004), and the US Army Research Office (W911NF-17-1-0448).



\bibliographystyle{IEEEtran}
\bibliography{infinity-mirror}

\begin{thebibliography}{10}
\providecommand{\url}[1]{#1}
\csname url@samestyle\endcsname
\providecommand{\newblock}{\relax}
\providecommand{\bibinfo}[2]{#2}
\providecommand{\BIBentrySTDinterwordspacing}{\spaceskip=0pt\relax}
\providecommand{\BIBentryALTinterwordstretchfactor}{4}
\providecommand{\BIBentryALTinterwordspacing}{\spaceskip=\fontdimen2\font plus
\BIBentryALTinterwordstretchfactor\fontdimen3\font minus
  \fontdimen4\font\relax}
\providecommand{\BIBforeignlanguage}[2]{{%
\expandafter\ifx\csname l@#1\endcsname\relax
\typeout{** WARNING: IEEEtran.bst: No hyphenation pattern has been}%
\typeout{** loaded for the language `#1'. Using the pattern for}%
\typeout{** the default language instead.}%
\else
\language=\csname l@#1\endcsname
\fi
#2}}
\providecommand{\BIBdecl}{\relax}
\BIBdecl

\bibitem{erdos1960evolution}
P.~Erd{\H{o}}s and A.~R{\'e}nyi, ``On the evolution of random graphs,''
  \emph{Publ. Math. Inst. Hung. Acad. Sci}, vol.~5, no.~1, pp. 17--60, 1960.

\bibitem{chung2002average}
F.~Chung and L.~Lu, ``The average distances in random graphs with given
  expected degrees,'' \emph{Proceedings of the National Academy of Sciences},
  vol.~99, no.~25, pp. 15\,879--15\,882, 2002.

\bibitem{watts1998collective}
D.~J. Watts and S.~H. Strogatz, ``Collective dynamics of
  ‘small-world’networks,'' \emph{nature}, vol. 393, no. 6684, pp. 440--442,
  1998.

\bibitem{barabasi1999emergence}
A.-L. Barab{\'a}si and R.~Albert, ``Emergence of scaling in random networks,''
  \emph{science}, vol. 286, no. 5439, pp. 509--512, 1999.

\bibitem{bianconi2001competition}
G.~Bianconi and A.-L. Barab{\'a}si, ``Competition and multiscaling in evolving
  networks,'' \emph{EPL (Europhysics Letters)}, vol.~54, no.~4, p. 436, 2001.

\bibitem{ravasz2003hierarchical}
E.~Ravasz and A.-L. Barab{\'a}si, ``Hierarchical organization in complex
  networks,'' \emph{Phys. Rev. E}, vol.~67, no.~2, p. 026112, 2003.

\bibitem{lancichinetti2008benchmark}
A.~Lancichinetti, S.~Fortunato, and F.~Radicchi, ``Benchmark graphs for testing
  community detection algorithms,'' \emph{Physical review E}, vol.~78, no.~4,
  p. 046110, 2008.

\bibitem{pfeiffer2012fast}
J.~J. Pfeiffer, T.~La~Fond, S.~Moreno, and J.~Neville, ``Fast generation of
  large scale social networks while incorporating transitive closures,'' in
  \emph{SocialCom Workshop on Privacy, Security, Risk and Trust
  (PASSAT)}.\hskip 1em plus 0.5em minus 0.4em\relax IEEE, 2012, pp. 154--165.

\bibitem{mussmann2014assortativity}
S.~Mussmann, J.~Moore, J.~J. Pfeiffer, and J.~Neville~III, ``Assortativity in
  chung lu random graph models,'' in \emph{Workshop on Social Network Mining
  and Analysis}.\hskip 1em plus 0.5em minus 0.4em\relax ACM, 2014, p.~3.

\bibitem{mussmann2015incorporating}
S.~Mussmann, J.~Moore, J.~J. Pfeiffer~III, and J.~Neville, ``Incorporating
  assortativity and degree dependence into scalable network models.'' in
  \emph{AAAI}, 2015, pp. 238--246.

\bibitem{kolda2014scalable}
T.~G. Kolda, A.~Pinar, T.~Plantenga, and C.~Seshadhri, ``A scalable generative
  graph model with community structure,'' \emph{SIAM Journal on Scientific
  Computing}, vol.~36, no.~5, pp. C424--C452, 2014.

\bibitem{karrer2011stochastic}
B.~Karrer and M.~E. Newman, ``Stochastic blockmodels and community structure in
  networks,'' \emph{Physical review E}, vol.~83, no.~1, p. 016107, 2011.

\bibitem{funke2019stochastic}
T.~Funke and T.~Becker, ``Stochastic block models: A comparison of variants and
  inference methods,'' \emph{PloS one}, vol.~14, no.~4, 2019.

\bibitem{aicher2013adapting}
C.~Aicher, A.~Z. Jacobs, and A.~Clauset, ``Adapting the stochastic block model
  to edge-weighted networks,'' \emph{arXiv preprint arXiv:1305.5782}, 2013.

\bibitem{larremore2014efficiently}
D.~B. Larremore, A.~Clauset, and A.~Z. Jacobs, ``Efficiently inferring
  community structure in bipartite networks,'' \emph{Physical Review E},
  vol.~90, no.~1, p. 012805, 2014.

\bibitem{peixoto2015inferring}
T.~P. Peixoto, ``Inferring the mesoscale structure of layered, edge-valued, and
  time-varying networks,'' \emph{Physical Review E}, vol.~92, no.~4, p. 042807,
  2015.

\bibitem{peixoto2014hierarchical}
------, ``Hierarchical block structures and high-resolution model selection in
  large networks,'' \emph{Physical Review X}, vol.~4, no.~1, p. 011047, 2014.

\bibitem{robins2007introduction}
G.~Robins, P.~Pattison, Y.~Kalish, and D.~Lusher, ``An introduction to
  exponential random graph (p*) models for social networks,'' \emph{Social
  networks}, vol.~29, no.~2, pp. 173--191, 2007.

\bibitem{leskovec2010kronecker}
J.~Leskovec, D.~Chakrabarti, J.~Kleinberg, C.~Faloutsos, and Z.~Ghahramani,
  ``Kronecker graphs: An approach to modeling networks,'' \emph{Journal of
  Machine Learning Research}, vol.~11, no. Feb, pp. 985--1042, 2010.

\bibitem{leskovec2007scalable}
J.~Leskovec and C.~Faloutsos, ``Scalable modeling of real graphs using
  kronecker multiplication,'' in \emph{ICML}.\hskip 1em plus 0.5em minus
  0.4em\relax ACM, 2007, pp. 497--504.

\bibitem{gleich2012moment}
D.~F. Gleich and A.~B. Owen, ``Moment-based estimation of stochastic kronecker
  graph parameters,'' \emph{Internet Mathematics}, vol.~8, no.~3, pp. 232--256,
  2012.

\bibitem{aguinaga2018learning}
S.~Aguinaga, D.~Chiang, and T.~Weninger, ``Learning hyperedge replacement
  grammars for graph generation,'' \emph{IEEE transactions on pattern analysis
  and machine intelligence}, vol.~41, no.~3, pp. 625--638, 2018.

\bibitem{sikdar2019modeling}
S.~Sikdar, J.~Hibshman, and T.~Weninger, ``Modeling graphs with vertex
  replacement grammars,'' in \emph{ICDM}.\hskip 1em plus 0.5em minus
  0.4em\relax IEEE, 2019.

\bibitem{hibshman2019towards}
J.~Hibshman, S.~Sikdar, and T.~Weninger, ``Towards interpretable graph modeling
  with vertex replacement grammars,'' in \emph{BigData}.\hskip 1em plus 0.5em
  minus 0.4em\relax IEEE, 2019.

\bibitem{you2018graphrnn}
J.~You, R.~Ying, X.~Ren, W.~L. Hamilton, and J.~Leskovec, ``Graphrnn:
  Generating realistic graphs with deep auto-regressive models,'' 2018.

\bibitem{salha2020simple}
G.~Salha, R.~Hennequin, and M.~Vazirgiannis, ``Simple and effective graph
  autoencoders with one-hop linear models,'' \emph{arXiv preprint
  arXiv:2001.07614}, 2020.

\bibitem{kipf2016variational}
T.~N. Kipf and M.~Welling, ``Variational graph auto-encoders,'' \emph{NIPS
  Workshop on Bayesian Deep Learning}, 2016.

\bibitem{li2018learning}
Y.~Li, O.~Vinyals, C.~Dyer, R.~Pascanu, and P.~Battaglia, ``Learning deep
  generative models of graphs,'' \emph{arXiv preprint arXiv:1803.03324}, 2018.

\bibitem{yun2019graph}
S.~Yun, M.~Jeong, R.~Kim, J.~Kang, and H.~J. Kim, ``Graph transformer
  networks,'' in \emph{Advances in Neural Information Processing Systems},
  2019, pp. 11\,960--11\,970.

\bibitem{bojchevski2018netgan}
A.~Bojchevski, O.~Shchur, D.~Z{\"u}gner, and S.~G{\"u}nnemann, ``Netgan:
  Generating graphs via random walks,'' \emph{arXiv preprint arXiv:1803.00816},
  2018.

\bibitem{koutra2013deltacon}
D.~Koutra, J.~T. Vogelstein, and C.~Faloutsos, ``Deltacon: A principled
  massive-graph similarity function,'' in \emph{Proceedings of the 2013 SIAM
  International Conference on Data Mining}.\hskip 1em plus 0.5em minus
  0.4em\relax SIAM, 2013, pp. 162--170.

\bibitem{koutra2016deltacon}
D.~Koutra, N.~Shah, J.~T. Vogelstein, B.~Gallagher, and C.~Faloutsos,
  ``Deltacon: principled massive-graph similarity function with attribution,''
  \emph{ACM Trans. on Knowledge Discovery from Data}, vol.~10, no.~3, p.~28,
  2016.

\bibitem{liu2018cut}
Q.~Liu, Z.~Dong, and E.~Wang, ``Cut based method for comparing complex
  networks,'' \emph{Scientific reports}, vol.~8, no.~1, pp. 1--11, 2018.

\bibitem{bagrow2019information}
J.~P. Bagrow and E.~M. Bollt, ``An information-theoretic, all-scales approach
  to comparing networks,'' \emph{Applied Network Science}, vol.~4, no.~1,
  p.~45, 2019.

\bibitem{wilson2008study}
R.~C. Wilson and P.~Zhu, ``A study of graph spectra for comparing graphs and
  trees,'' \emph{Pattern Recognition}, vol.~41, no.~9, pp. 2833--2841, 2008.

\bibitem{ahmed2015efficient}
N.~K. Ahmed, J.~Neville, R.~A. Rossi, and N.~Duffield, ``Efficient graphlet
  counting for large networks,'' in \emph{2015 IEEE International Conference on
  Data Mining}.\hskip 1em plus 0.5em minus 0.4em\relax IEEE, 2015, pp. 1--10.

\bibitem{prvzulj2007biological}
N.~Pr{\v{z}}ulj, ``Biological network comparison using graphlet degree
  distribution,'' \emph{Bioinformatics}, vol.~23, no.~2, pp. e177--e183, 2007.

\bibitem{aguinaga2016infinity}
S.~Aguinaga and T.~Weninger, ``The infinity mirror test for analyzing the
  robustness of graph generators,'' in \emph{KDD Workshop on Mining and
  Learning with Graphs}.\hskip 1em plus 0.5em minus 0.4em\relax ACM, 2016.

\bibitem{seshadhri2013depth}
C.~Seshadhri, A.~Pinar, and T.~G. Kolda, ``An in-depth analysis of stochastic
  kronecker graphs,'' \emph{Journal of the ACM (JACM)}, vol.~60, no.~2, pp.
  1--32, 2013.

\bibitem{dai2020scalable}
H.~Dai, A.~Nazi, Y.~Li, B.~Dai, and D.~Schuurmans, ``Scalable deep generative
  modeling for sparse graphs,'' in \emph{International Conference on Machine
  Learning}.\hskip 1em plus 0.5em minus 0.4em\relax PMLR, 2020, pp. 2302--2312.

\bibitem{seshadhri2012bter}
C.~Seshadhri, T.~G. Kolda, and A.~Pinar, ``Community structure and scale-free
  collections of erdos-renyi graphs,'' \emph{Phys. Rev. E}, vol.~85, p. 056109,
  May 2012.

\bibitem{karrer2011sbm}
B.~Karrer and M.~E.~J. Newman, ``Stochastic blockmodels and community structure
  in networks,'' \emph{Phys. Rev. E}, vol.~83, p. 016107, Jan 2011.

\bibitem{aguinaga2016growing}
S.~Aguinaga, R.~Palacios, D.~Chiang, and T.~Weninger, ``Growing graphs from
  hyperedge replacement grammars,'' in \emph{CIKM}.\hskip 1em plus 0.5em minus
  0.4em\relax ACM, 2016.

\bibitem{tantardini2019comparing}
M.~Tantardini, F.~Ieva, L.~Tajoli, and C.~Piccardi, ``Comparing methods for
  comparing networks,'' \emph{Scientific reports}, vol.~9, no.~1, pp. 1--19,
  2019.

\bibitem{bagrow2008portraits}
J.~P. Bagrow, E.~M. Bollt, J.~D. Skufca, and D.~Ben-Avraham, ``Portraits of
  complex networks,'' \emph{EPL (Europhysics Letters)}, vol.~81, no.~6, p.
  68004, 2008.

\bibitem{klimek2013triadic}
P.~Klimek and S.~Thurner, ``Triadic closure dynamics drives scaling laws in
  social multiplex networks,'' \emph{New Journal of Physics}, vol.~15, no.~6,
  p. 063008, 2013.

\end{thebibliography}


%
\begin{IEEEbiography}[{\includegraphics[width=1in,height=1.25in,clip,]{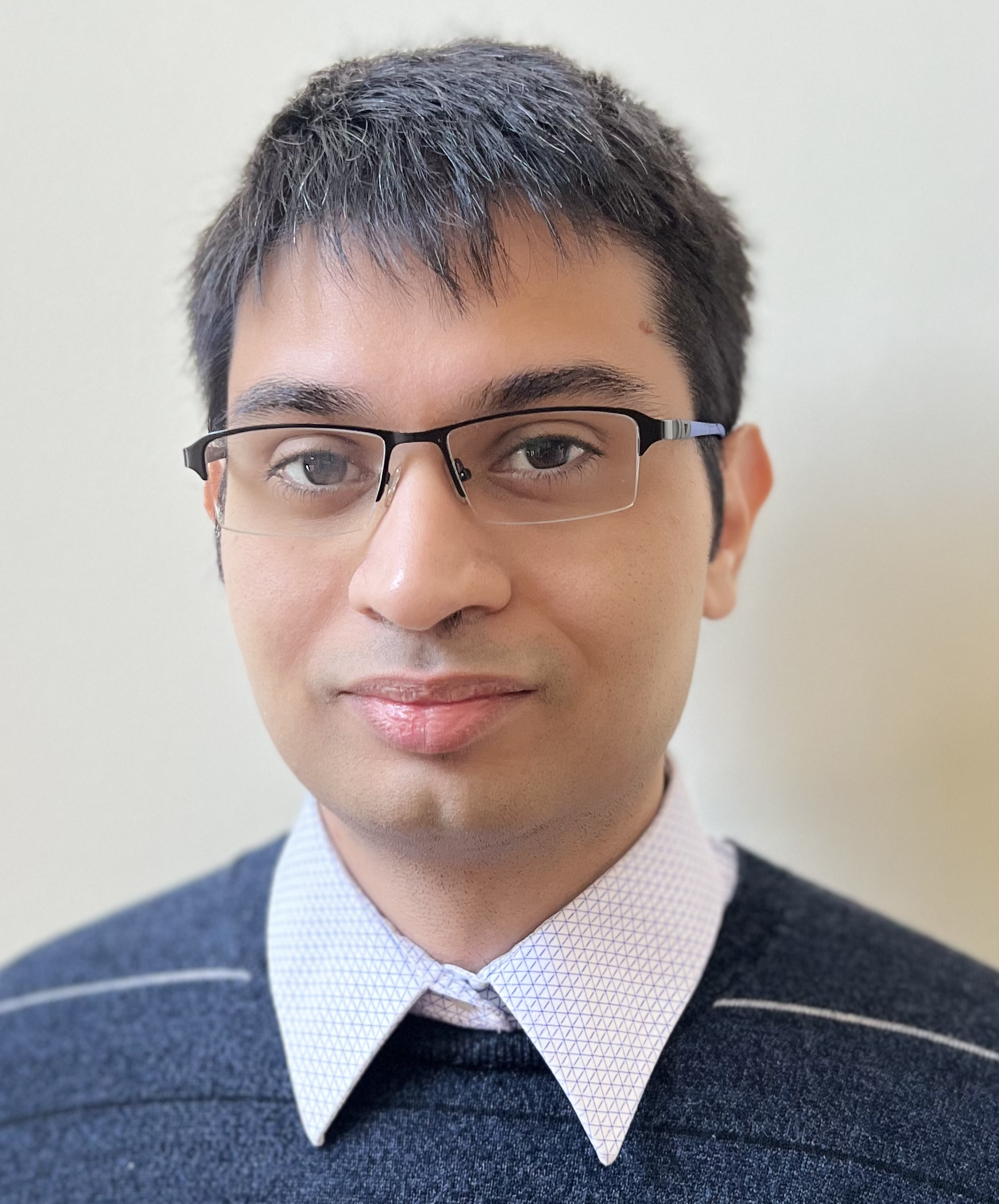}}]{Satyaki Sikdar}
Satyaki Sikdar received the Ph.D. degree from the University of Notre Dame, Notre Dame, IN, USA, in 2021. He will join the Luddy School of Informatics, Computing, and Engineering at Indiana University as a
Post-Doctoral Fellow in January 2022. His research is at the intersection of graph theory, formal language theory, and data mining. His work has been published in IEEE ICDM, IEEE BigData, ACM WSDM, and KAIS.
\end{IEEEbiography}

\begin{IEEEbiography}[{\includegraphics[width=1in,height=1.25in,clip,]{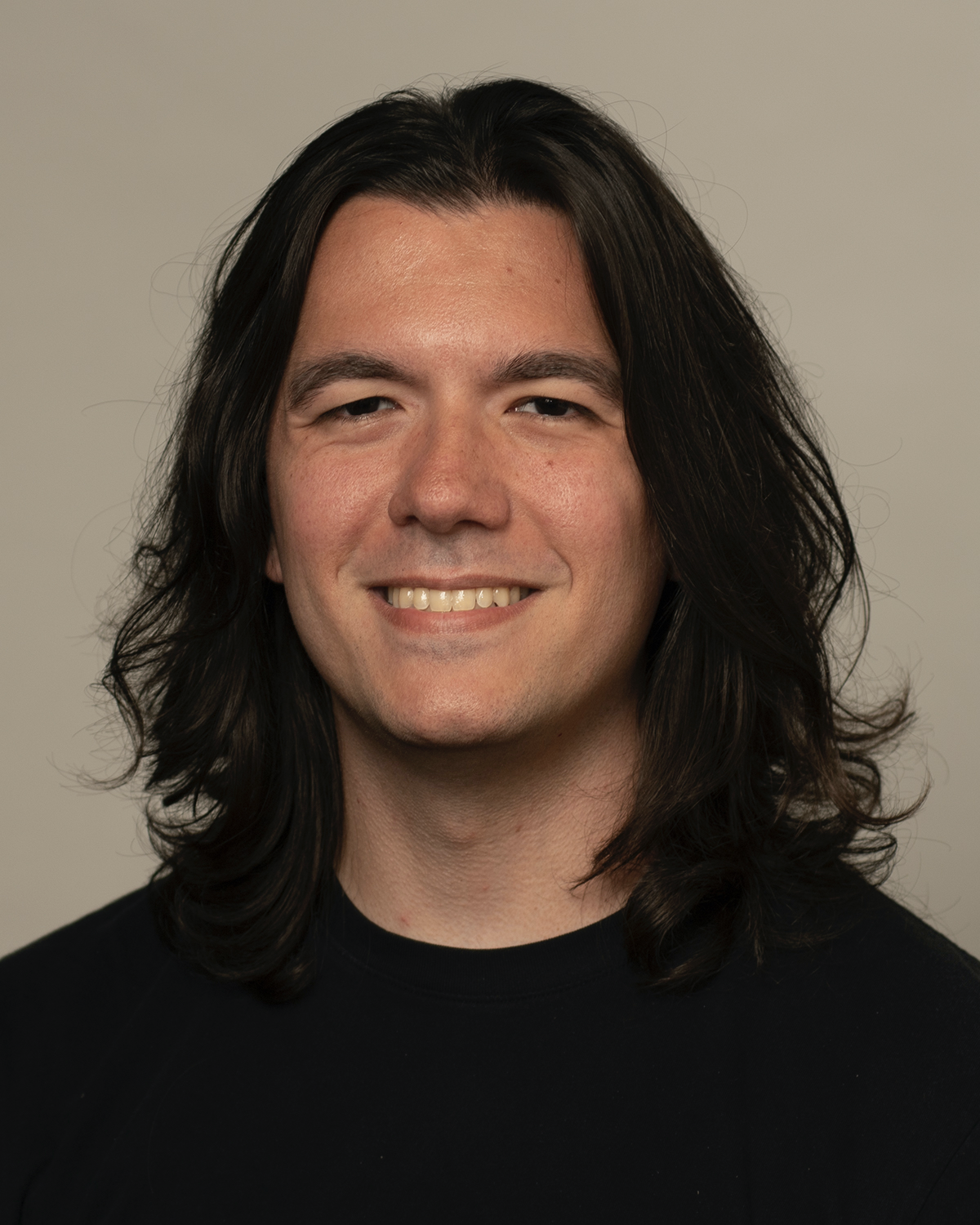}}]{Daniel Gonzalez Cedre}
Daniel Gonzalez Cedre received the M.S. degree in financial mathematics from the Department of Mathematics at Florida State University, Tallahassee, FL, USA, in 2019.
He is working towards the Ph.D. degree with the Department of Computer Science and Engineering at the University of Notre Dame, Notre Dame, IN, USA.
His current research interests are in dynamical graph systems and graph modeling, working towards a mathematical understanding of real-world complex networks.
\end{IEEEbiography}

\begin{IEEEbiography}[{\includegraphics[width=1in,height=1.25in,clip,]{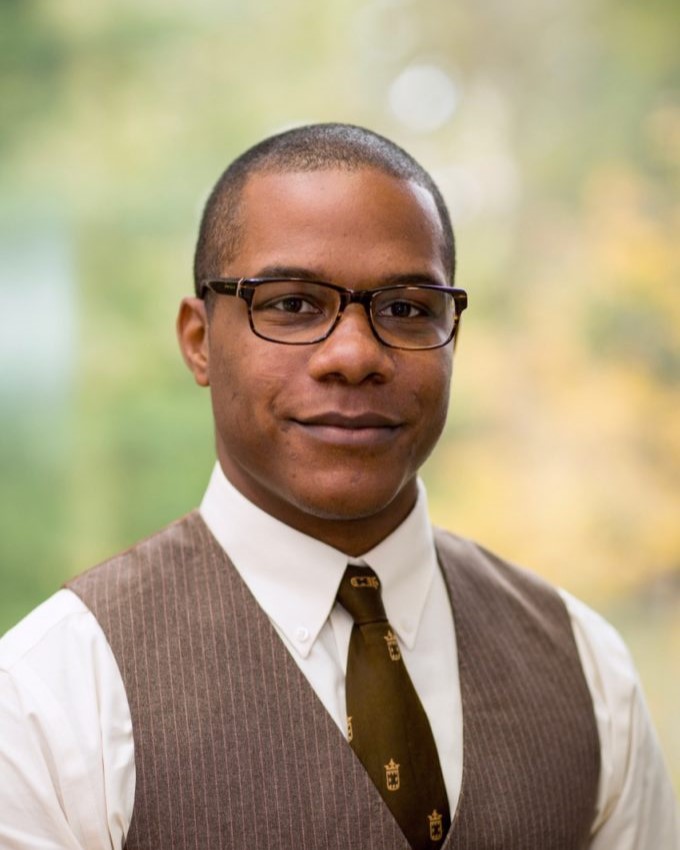}}]{Trenton W. Ford}
Trenton W. Ford received the M.S. degree in Computer Science \& Engineering from the University of Notre Dame, Notre Dame, IN, USA, in 2021. He is a Ph.D. Candidate in the Department of Computer Science and Engineering at the University of Notre Dame. He is also an Editorial Fellow with The Bulletin of the Atomic Scientists where he writes about disruptive technologies. His current research interests are in semantic analysis in social media systems. 
\end{IEEEbiography}

\begin{IEEEbiography}[{\includegraphics[width=1in,height=1.25in,clip,]{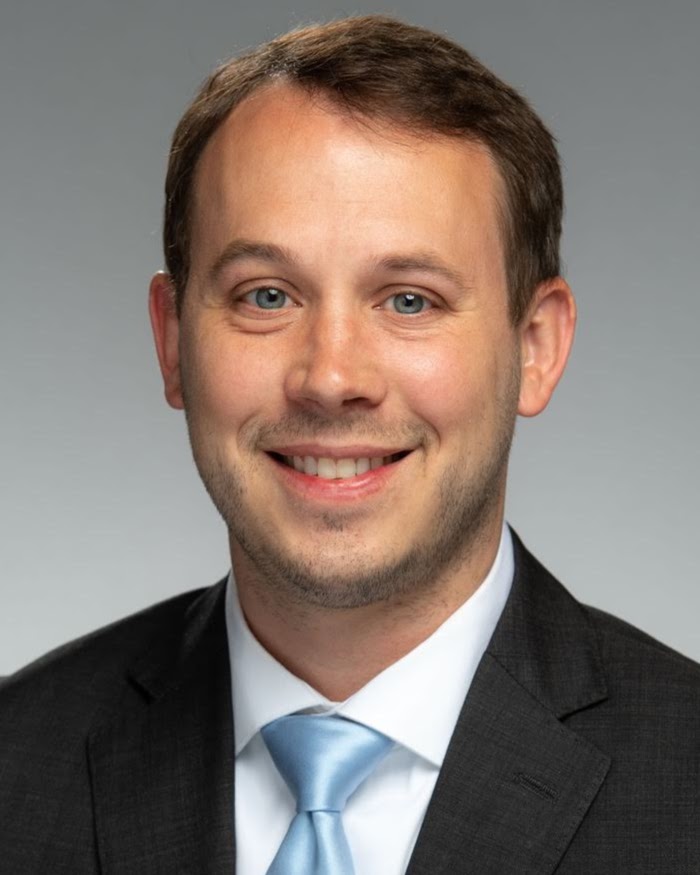}}]{Tim Weninger}
Tim Weninger received the Ph.D. degree from the University of Illinois Urbana–Champaign,
Champaign, IL, USA, in 2013. He is the Frank M. Friemann Collegiate Associate Professor of Engineering with the Department of Computer Science and Engineering at the University of Notre Dame, Notre Dame, IN, USA. His current research interests include the intersection of social media, data mining, and network science, in which he studies how humans create and consume networks of information. He has received research grants from NSF, AFOSR, ARO, USAID, DARPA, and the John Templeton Foundation.

\end{IEEEbiography}

\vfill







\end{document}